\documentstyle[aps,preprint]{revtex}
\input epsfig
\begin{document}
\draft
\newcommand{\Bl}{ \left( }
\newcommand{\Br}{ \right)}
\title{The approach to equilibrium in $N$-body gravitational systems}

\author{A.A. El-Zant} 
\address{Astronomy Centre, University of Sussex,
Brighton BN1 9QH} 

\address{Physics Department, Technion --- 
Israel Institute of Technology, Haifa 32000, Israel (present address)}

\date{\today}
\maketitle

\begin{abstract}
The evolution of closed gravitational systems is studied by
means of $N$-body simulations. This, as well as being interesting in its own 
right, provides insight into the dynamical and statistical mechanical properties 
of gravitational systems: the possibility of the existence of stable 
equilibrium states and the associated relaxation time would provide an ideal
situation where relaxation theory can be tested. Indeed, these states are
found to exist for {\em single}  mass $N$-body systems, and
the condition  condition for this is simply that
obtained  from elementary thermodynamical 
considerations applied to self-gravitating ideal gas spheres.
However, even when this condition is satisfied, some initial states may
not end as isothermal spheres. It is therefore only a necessary condition. 
Simple considerations also
predict that, for fixed total mass, energy and radius, stable isothermal
spheres are unique. Therefore, statistically irreversible 
perturbations to the
density profile caused by the accumulation of massive particles near the
centre of multimass systems, destroy these equilibria if the aforementioned
quantities are kept fixed. The time-scale for this to happen was found to be
remarkably short (a few dynamical times when $N= 2500$) in systems undergoing
violent relaxation. 
The time taken to achieve thermal equilibrium depended 
on the initial conditions and 
could be comparable to a dynamical time (even when the conditions for violent
relaxation were not satisfied) or the two body relaxation time. The relaxation
time for velocity anisotropies was intermediate between these two time-scales, 
being long compared to the dynamical time but much (about four times) shorter 
than the time-scale of energy relaxation. This last result, 
along with the observation
of the anomalously rapid mass segregation in some situations, suggests that
in gravitational systems, different quantities may relax at different rates,
and that the thermal (two body) relaxation time-scale, 
even if accurate for energy relaxation of single mass systems,
may not be universal. 
This in turn indicates  that the issue of relaxation in gravitational
systems is far from being a closed subject.

\end{abstract}


\section{Introduction and motivation}
\label{sphex:genev}
 
The peculiarity of gravitational interactions compared to those governing 
laboratory systems can be illustrated by the following  example. 
An orbiting satellite
on a circular path loses energy (by dissipation say) 
and drops to another (also circular)
orbit  with a smaller radius. The result is that its circular 
velocity {\em increases},
even though its total energy decreased. This situation is easily translated 
into the 
language of large $N$-body systems in virial equilibrium. 
In this case, the kinetic and
potential energies are related by 
$2T + V = 0$,
which implies  that the total
energy $E=-T$. Decreasing the
total energy will therefore be equivalent to  increasing the kinetic 
energy~\protect\cite{BT,Saslaw,Chandra}. 
That is, the system
has {\em negative specific heat} --- since taking energy  away from it heats 
it up.
If one is able to apply  this logic to  subsystems of a gravitational 
configuration in dynamical 
equilibrium, one can see that ``heat''  will  
{\em effectively} flow from hotter 
regions (those having larger average random kinetic energy) to 
cooler ones --- which implies that any 
temperature gradient is enhanced instead of being erased.

Any increase in  the kinetic energy of a system in virial equilibrium 
 is bound  (by the virial relation) 
to decrease its potential energy.
For, in a spherical  system of mass $M$ and radius $R$ the potential energy 
is given by
\begin{equation}
V = - \frac{\gamma G M^{2}}{R},  
\end{equation}
where $\gamma$ is a constant which depends on the density distribution 
of the system 
(and is larger
for centrally concentrated objects).
Decreasing the potential energy will therefore imply a decrease in 
radius or an 
increase in central concentration. 
If  a situation arises where a gravitational system can, to a first
approximation, be considered as to be composed of 
a centrally concentrated core  
in virial equilibrium which is hotter than a surrounding shell, 
 the core will lose energy to the surrounding  shell and contract, while
getting still hotter. If the surrounding halo cannot heat fast enough, we
 have a runaway instability, with the
 system evolving towards states that are less and less homogeneous in 
both the  
physical and velocity spaces~\protect\cite{BT,Saslaw}. 
This phenomenon has been termed a 
``gravothermal catastrophe'' by Lynden Bell and 
Wood~\protect\cite{Lynden Bell and Wood}.
 A system undergoing such evolution, instead of tending towards a 
most probable final equilibrium state,
can increase its Boltzmann entropy 
indefinitely --- simply by evolving a denser core 
and a more diffuse halo ~\protect\cite{BT}.

Thermal equilibrium configurations do nevertheless exist
for gravitational systems. For open (spatially unbound) systems 
these however turn out to be infinite.
In a closed spherical system, the boundary has the effect of adding an 
external pressure term to the virial equation. 
The kinetic and potential energies are then related by $2 T/V \geq 1$. 
If  this ratio is very large, the
behaviour will be similar to that of a normal laboratory system --- and
a constant  temperature (almost) spatially homogeneous 
equilibrium configuration will exist. 
If, on the other hand, the potential energy is comparable to the kinetic 
energy, then
the peculiar effects related to the nature of  gravitational interactions
will dominate. It turns out that the precise criterion for avoiding 
 gravothermal instability  in an {\em ideal gas} sphere 
is~\protect\cite{Lynden Bell and Wood,Antonov,Padmanabhan}
\begin{equation}
\mu = \frac{E R}{G M^{2}} \geq - 0.335 .
\label{sphex:gravo}
\end{equation}

Enclosed gas spheres for which~(\ref{sphex:gravo}) 
 holds, may reach constant temperature  
equilibria with a well defined 
asymptotic statistical distribution. These are often 
referred to as (Lane-Embden) isothermal  spheres, because their 
 final  distribution functions obey the usual 
Maxwell-Boltzmann statistics of an isothermal gas.
It is easy to see that for systems 
in virial equilibrium $\frac{E R}{G M^{2}} = - \gamma/2$.
In the  case when the density
decreases with radius, $\gamma$ cannot be smaller than $3/5$ . 
Almost  all systems in virial equilibrium therefore do not
tend towards thermal equilibrium  states
 --- since  dynamical equilibria of self-gravitating 
are necessarily  strongly inhomogeneous, with density 
decreasing with radius. 
In such systems  the central 
core may behave almost as if
it is an independent 
subsystem in virial equilibrium  and  
its (rather weak but important) interaction with the outer regions will 
lead to a gravothermal catastrophe.

Maximum entropy solutions of the Collisionless Boltmann Equation (CBE)
which have constant velocity dispersions also
obey Maxwell-Bolztamnn statistics and, by analogy, are also referred 
to as isothermal spheres.
Intuitively, therefore,  one would expect condition~(\ref{sphex:gravo})
 to  hold for spherical stellar systems~\protect\cite{BT}.
Indeed, {\em local} stability analysis based on detailed treatment 
of the statistical mechanics
of gravitating systems seems to 
confirm this~\protect\cite{Horowitz and Katz,Cally and Monaghan}.
However, the aforementioned solutions of the CBE are obtained by 
constraining the energy, 
not the temperature. Therefore, strictly speaking,
 they represent constant-energy solutions, rather than 
isothermal solutions in the standard sense.

By global minimisation of  free energy 
one can obtain truly isothermal solutions~\cite{Kiessling}.
In this case however,
the  conditions for the 
existence of stable isothermal
equilibrium are found to be much stricter. For an 
unsoftened $1/r$ potential, an isothermal
gravitational system which obeys classical mechanics will contract 
to a point.
For softened systems (with which we shall be concerned here),
 other isothermal solutions
exist. When $\mu$ is very large (of the order of the inverse
 of the softening length),
these overlap with the Lane-Emden isothermal spheres. 
For smaller values of $\mu$
the latter only represent {\em local} entropy maxima. 
Thus, as shown by Kiessling~\protect\cite{Kiessling},
 even systems satisfying~(\ref{sphex:gravo}) may undergo 
gravothermal catastrophe. In this case, Lane-Embden ``isothermal 
spheres'' no longer represent true isothermal equilibria.
To distinguish them from the latter, such states will, in general, be 
referred to as {\em thermal} equilibria (or isothermal spheres).
Other authors have doubted the very validity of the use of 
entropy maximisation or canonical ensemble statistical mechanics 
(on which all of the above conclusions are based)
in systems with long range forces~\protect\cite{Ipser}. 
The validity of relation~(\protect\ref{sphex:gravo}) will be one of the 
issues we will be examining  numerically in this paper.

The existence of a final equilibrium state provides an idealised situation
 whereas 
the dynamics of gravitational systems can be studied for long times 
by means of particle simulations, without the macroscopic time dependence  
necesarilly  characteristic of the evolution of open gravitational systems. 
This may for example be useful
in studying the stability of trajectories of particles in $N$-body systems, 
their ergodic properties, the diffusion rates of their action variables 
and the possible
relationship between these properties 
(for more details concerning the motivations
behind such an exercise see 
the concluding section of Ref.~\protect\cite{El-Zant2}). 
So far this has been done
only for 1-dimensional gravitational systems~\protect\cite{Tsuchiya et al.}. 
It is not clear however how the behaviour of 
 these systems, consisting of a series of infinite sheets, relates 
to that of generic 3-dimensional 
gravitational ones. In
particular, the force between the ``particles'' in these systems is constant
 most of the time but
discontinuous when  they cross each other, thus much of the body 
of rigorous results concerning the stability of dynamical systems, which is evidently 
useful in  understanding the properties mentioned above, does not 
apply~\protect\cite{Pesin}. Nevertheless, we will find some interesting parallels
between the behaviour of these systems and the 3-dimensional softened systems 
described here. Comparison between the behaviour of the two types of systems may thus 
guide further exploration of relevant properties of  1-d systems (which may 
be simulated more accurately for much longer times). 

The existence of a 
well defined relaxation time  provides a clean test for the 
relaxation theory 
of gravitational systems and insight into 
the nature of the relaxation process. 
This is important 
for the following reasons. First there is still much debate as to how a 
system  actually achieves a  relaxed {\em dynamical equilibrium}. 
Violent relaxation, the original mechanism
suggested for this process, is still not very well 
understood~\protect\cite{Saslaw,Kandrupcr}. 
It is still not clear whether
a dynamically relaxed final state is necessarily related to this process 
or whether 
the conditions for violent relaxation did exist in the early history of most 
stellar systems. The second reason is related to the question as to under 
which conditions (if any)
classical ``collisional'' relaxation 
theory~\protect\cite{BT,Saslaw,Spitzer,Rev Mods,Heggie and Meylan} holds. 
Controversial since its original
formulation~\protect\cite{Kurth} due to its description of discreteness 
effects as independent two body
encounters added to the mean field motion 
(something that cannot be a priori justified
in a non-linear system), it is nevertheless commonly accepted as valid. 
Although there seems
to be some justification to its use in describing the energy relaxation 
of particles in
$N$-body gravitational systems~\protect\cite{El-Zant1},
and indeed much of the limited numerical evidence that there 
is~\protect\cite{Theuns,Giersz and Heggie,Farouki and Salpeter1,Farouki and Salpeter2}
seems to point in 
that direction, it seems  hardly justified that the validity of this 
time-scale be taken
for granted (as is almost always done in research on
dynamical astronomy) without strict tests, pending
 rigorous theoretical justifications for the approximations
made in deriving it~\protect\cite{Comtrem}. Indeed, it is not uncommon 
to find in numerical simulations that
 significant effects arising from discreteness noise  
take place on a scale much shorter than the standard 
two body relaxation time~\protect\cite{Come1}.

Whether it does predict the energy relaxation time correctly or not, 
the dynamical 
picture upon which standard relaxation theory is based appears to be flawed, 
since it
implicitly assumes that gravitational $N$-body systems are  
integrable and
remain so under perturbations due to discreteness. 
This is modelled as {\em additive} stochastic noise. The final result 
being a simple linear superposition of two independent solutions ---
 the regular motion in the mean field 
and that under the influence of the stochastic force.
However, it is now well 
known~\protect\cite{Miller,Goodman et al.,Kandrup et al.} 
 that large $N$-Body gravitational systems display sensitivity to changes 
in their initial
conditions which are characteristic of chaotic dynamical systems,
 and that this appears
to be related to their being systems with predominantly negative 
configuration space 
curvature~\protect\cite{Gurzadyans,Kandrup1,El-Zant1,El-Zant2,szyi}
 --- thus having qualitative properties very different from those of 
integrable systems. 
It is therefore plausible that quantities that depend on the details of the 
$N$-body trajectories
(as opposed to quantities like energy which is a scalar 
path independent integral of motion) 
 may relax on time-scales that are different from the standard 
relaxation 
time~\protect\cite{elmore}. 
This may affect important observable quantities, like the degree anisotropy 
in a given 
system for example. In addition, there are situations when it is 
clear that the relaxation
phenomena involved are beyond the applicability of simple classical two
 body relaxation theory.
Examples of such effects include 
 those arising from the motion of massive particles
in a $N$-body system~\protect\cite{Saslaw} and the interaction of 
discreteness noise with the global mean field modes of a 
system~\cite{Zawin}.
Theories concerning such effects are far less well established
 than standard two body relaxation
(whereas the approximations made, although not resting on rigorous 
theoretical grounds, are at
least familiar from  the  theory  of stochastic processes and well 
formulated~\protect\cite{Chandra2}).

This paper has the basic aim of testing, through direct $N$-body simulations, 
some of the theory concerning isothermal spheres, and showing how
conclusions of important physical interest --- mainly
concerning the dynamical  relaxation  of gravitational systems --- 
can be obtained from this type of study. 
Thus we show the existence  of stable 
$N$-body realisations of isothermal spheres for single mass systems and 
determine the characteristic
time-scales of achieving this state (Section~\ref{slowap}).
We also examine the time-scale of relaxation
towards isotropy for systems with initialy anisotropic velocity dispersion 
tensor (Section~\ref{anis}). 
By simulating multi-mass systems where heavier particles tend
to reside towards the centre --- thus modifying the isothermal sphere 
density profile ---
we check if, for a given total energy, mass and radius, such configurations 
have a unique density distribution (Section~\ref{multi}).
 In the process, the time-scale  of mass segregation is evaluated. 
Finally (Section~\ref{phatran}), 
we will check if the condition distinguishing $N$-body
 systems that collapse from ones that find a stable thermal 
equilibrium state 
is similar
to that for  the existence of a stable self-gravitating ideal 
gas sphere~(Eq.~\ref{sphex:gravo}).
We start by describing the direct summation gravitational 
$N$-body code used in this paper, 
and  other technical details like 
the generation of initial data and the units 
used~(Sections~\ref{sphex:NBODY2} and~\ref{sphex:closed}).

\section{Numerical method}
\label{sphex:NBODY2}

The main computational load in the integration of a gravitational 
$N$-body system lies in 
the calculation of the particle-particle forces, which, when summed directly,
take a time proportional to the square of the particle number to compute.
Many different techniques have 
been devised in order to speed up the force calculations
 (for example  ``tree'' techniques ~\protect\cite{Barnes and Hut} 
and adaptive particle mesh methods~\protect\cite{Couchman et al.}).
As usual, there is always a tradeoff between computational efficiency 
and accuracy.
While the aforementioned techniques are very powerful (the CPU time spent in 
the force calculations scales as $N \log N$ at the worst), 
they are not very accurate 
--- one does not expect that, in general, the  particle-particle interactions
 will be calculated
to an accuracy much better than few percent. 
This makes them unsuitable for work 
in which high
accuracy is required, or when the nature of the motion under the influence 
of gravitational forces is itself the object of study.

The NBODY2
 code~\cite{Aarseth} used to run the simulations described in this paper is one 
of the many 
efficient  routines devised by Sverre Aarseth and which are 
kindly provided
by him upon request. It is a direct summation code which uses individual 
time-steps for each particle
in the simulation~\protect\cite{Aarsetho} and speeds up the force 
calculation by using the
Ahmad-Cohen~\protect\cite{Ahmad and Cohen} neighbour scheme which, 
in the spirit of tree techniques, separates
 the force calculations for neighbouring particles and those further off ---
albeit in a somewhat different manner than tree methods. 
These improvements take into
 account the very different natural times ($\sim 1/\sqrt{\rho}$, $\rho$
being the local density) in a gravitational system and the fact that, at
a given point, the {\em irregular} force due 
to nearby neighbours varies much faster than the 
{\em regular} force due to particles further off.

The errors in the calculations (as measured by energy conservation)
are controlled by an
accuracy parameter $\eta$ which determines the size of the integration 
time-steps. 
These errors  are constant for values of $\eta$ below 
0.01 and increase 
 as $\eta^{2}$ for higher values~\protect\cite{Makino}. We have found that
a value of $\eta_{irr}=0.02$ (controlling the irregular time-step)
 gave reasonably accurate results 
while maintaining  efficient running of the code (both of these aspects  
depended on the type of enclosure bounding our systems as we will see below).
The tolerance 
parameter for the regular time-step was taken as $\eta_{reg}=0.04$. 
The softening length was fixed at $\epsilon_{p}=1/500$ .

To do any experiments on closed systems, 
we obviously have to find a practical numerical 
procedure for enclosing them. There is of course more than one way of doing 
this.
For example, one can just reverse the radial component of the 
velocities of particles
that are found to be beyond a certain radius. Alternatively, 
one can impose ``periodic
boundary conditions'', a situation in which particles escaping from one 
side  of a system
reappear on the opposite side. 
Both these conditions however destroy the smoothness of the
dynamical system under consideration --- in the first case the force 
can become infinite,
while in the second, some of the variables become discontinuous. 
It is upon the assumption
of smoothness that  many  of the rigorous results of dynamical systems
 stability theory 
and the conclusions drawn from it are based~\protect\cite{Pesin}. 
These may be important 
in understanding some properties of gravitational dynamics, 
since it appears that 
large gravitational systems are close to smooth hyperbolic ones 
and it has been suggested that this 
property may play a role in determining some of their dynamical 
characteristics~\protect\cite{Gurzadyans,Kandrup1,El-Zant1,El-Zant2,szyi}.

In order to conserve  the differentiability of (softened) 
$N$-body gravitational dynamical systems,
we have opted for the following procedure. 
We surround the system
by an ``elastic shell'', in the sense that a particle venturing 
beyond a given radius 
$r = r_{0}$ experiences a restoring force
\begin{equation}
{\bf F}_{res} = - m K (r - r_{0})^{2n -1} \hat{\bf r},
\label{sphex:lasforce}
\end{equation}
where $\hat{\bf r}$ is 
the radial unit vector at the particle's position and $m$ is its mass.
This central force law ensures the conservation of both 
the energy and angular momentum
of individual particles --- an important property if one is studying 
relaxational
phenomena.

Although the choice of the constants $K$ and $n$ is to a large 
extent arbitrary, $K$ has 
to be chosen so that particles do not venture too far beyond $r = r_{0}$. 
Therefore, for large excursions,  the force has to be strong.
However,
if the force rises too steeply
 at small excursions beyond $r_{0}$,  large errors
in the energies of the particles can result 
(as these gain or lose energy during their entries into and exits from the 
$r > r_{0}$ region when their velocities is largest).  
After a few trials, values of $K = 1300$ and $n=4$ were adopted. 
This ensured that, for the accuracy parameters adopted (see above),
the total energy change over a hundred
crossing times was always less than $1.5 \%$ and that particles almost
never ventured beyond 
$r - r_{0}=0.2$ and very rarely beyond $r - r_{0} =0.1$. The choice also
 ensured that the
inclusion of the boundary force did not slow down the computation too much
(the system of units used is described below).

\section{Initial conditions and related parameters}
\label{sphex:closed}

Throughout this paper we use the units of Heggie and 
Mathieu~\cite{Heggie and Mathieu}
 whereas the
total mass and the gravitational constant are set to unity. Except
 for some of the runs in Section~\ref{phatran} 
 where the initial density decreases according to a power law in 
the radius, 
we start our simulations from homogeneous spatial initial conditions
with the boundary at $r_0=1$, thus
fixing the potential energy at about $3/5$ 
(give or take effects due to particle noise and softening). 
The total energy is then  determined
by the initial virial ratio ($T/V$). If this is equal to 0.5 
(viral equilibrium in the 
absence of enclosure) then the {\em mean crossing time} is calculated
 from~\protect\cite{Aarseth} 
\begin{equation}
T_{cr}= M^{5/2}/(-2 E)^{3/2},
\label{sphex:crossing}
\end{equation}
which amounts to two time units. If the $T/V > 0.5$, the crossing time is 
shorter. However, considering that the crossing time is only defined as 
an order 
of magnitude quantity and that our systems will usually have virial 
ratios not too different from 0.5, we stick to this definition.

Systems were started from several initial velocity profiles.  
In the  more frequently used distributions, the velocity vectors take random 
directions with their magnitudes either decreasing or increasing
  with radius according to the
exponential law
\begin{equation}
V = V_{0} e^{- p r},
\label{sphex:VAI}
\end{equation}
where the central velocity is fixed by the chosen virial ratio
and radial density distribution.
 The parameter $p$ either takes a value $p=1$, in which case the 
velocity decreases towards the outside, or $p=-1$, 
for systems started with a  ``temperature inversion'', so that the velocity 
at the
edge of the system was $e$ times the velocity at the centre. 
We also examined the 
evolution of systems starting from anisotropic initial conditions  
with the  anisotropy
increasing towards the outside 
 according to the prescription
\begin{equation}
V_{l} = V_{l0} e^{- s \sqrt{l} r},
\label{sphex:vani}
\end{equation}
where $l=1,3$ denotes the Cartesian coordinates $x$, $y$ and $z$
 respectively. The
factor $s$ takes a value of $s=1$ unless otherwise stated. 
In one of the runs (Section~\ref{anis}) the system is started with 
two of the velocity coordinates (for all particles) set to zero.

In some of the runs it was important that our systems did not 
start too far from virial equilibrium, so that violent relaxation  
is (presumably) not effective. 
There is a well known formula (Ref. \cite{BT} Eq. 8P-2) 
for the condition of virial equilibrium
of a system on which external forces are applied.
Unfortunately however,
when the contribution of the external potential is included 
in the calculation of the virial ratio, 
this quantity 
is highly fluctuating (due to the large contribution from only a few 
particles which are 
beyond $r = r_{0}$). Systems starting from virial equilibrium including 
the boundary
force did not conserve that equilibrium. However, after looking at the long 
term evolution of the 
ratio $vir=2T/V$ in systems started this way, it was 
found that, for systems with initially homogeneous density,
 this quantity settled to a value between
$vir=1.32$ and $vir=1.38$ . 
In actual trials it was found that  isotropic
systems started from $vir=1.38$ and from homogeneous density states, 
 conserved this quantity to high accuracy 
(better than $3 \%$)~\cite{Note1}.
In the anisotropic case described by Eq.~(\ref{sphex:vani}) $vir$ was 
conserved to
better than $6 \%$ during
the evolution.

For all systems for which the initial value of $vir=1.38$, the  total
energy  is $E= - 0.185$ (this value includes the effect of a 
softening parameter in the Newtonian potential on the total energy). 
Therefore,
according to relation~(\ref{sphex:gravo}), these systems 
should evolve towards stable isothermal sphere configurations.
$vir$ will take this initial value (corresponding to a virial ratio of 0.69)
for  all the runs studied here, except for some 
of  those 
in Section~\ref{phatran}, where we vary the energy 
by changing the virial ratio, in an attempt to 
examine the validity of (\ref{sphex:gravo}).

To quantify the departure from thermal equilibrium, we either divide the 
region where 
$r \leq r_{0} + 0.1$
into  10 cells --- with the central cell enclosing a radius which is twice
the thickness of the surrounding shells --- 
and  calculate the trace of the 
velocity dispersion tensor at time $t$
 $\sigma_{t}$ in each of these 
cells. Or, alternatively, we divide the particles into ten groups 
depending on their distance from the centre.
Thus, the first set would contain the $N/10$ particles closest to 
the origin, the second set
contains the following $N/10$ particles as one moves outwards, and so on. 
The second method has the
advantage that the sets have equal numbers of particles, 
while the first procedure has (except for the inner cell) fixed
spatial resolution and therefore ensures that there are no large velocity  
gradients within the individual cells.
In both cases we calculate the RMS dispersion of the trace  
of the {\em velocity} 
dispersion tensor
from the relation
\begin{equation}
\sigma_{d}=\frac{\sqrt{\frac{1}{10}
\Sigma_{l=1}^{10}(\sigma^{l}_{t}-\bar{\sigma}_{t})^{2}}}{\bar{\sigma}_{t}}, 
\label{disperse}
\end{equation}
where a bar denotes an average over the cells. Usually the value of 
$\sigma_{d}$
using the two different methods of binning the data agreed reasonably 
well (within $10 - 20 \%$).
In what follows when we speak about the 
``dispersion'' or about $\sigma_{d}$ we will usually
 mean the average of $\sigma_{d}$  calculated
using the two different binning procedures described above.

\section{The approach to thermal equilibrium}
\label{slowap}

In this section we describe the evolution towards thermal equilibrium
of systems consisting of 2500 equal mass particles starting 
from some of the initial conditions described in the previous section.
Our object is to examine the existence of stable $N$-body isothermal spheres 
when simple thermodynamical theory 
predicts their existence 
 and to study the time-scale for 
achieving such equilibria when they exist.
The choice of the number of particles
was a compromise between the divergent requirements of studying systems  
in which the relaxation times towards dynamical and thermal
equilibrium should
 be significantly different (at least according to standard 
relaxation theory), and the large computing resources needed
by direct summation codes for high $N$. 

\subsection{The case when an intermediate dynamical equilibrium is reached}

\begin{figure}
\begin{center}
\epsfig{file=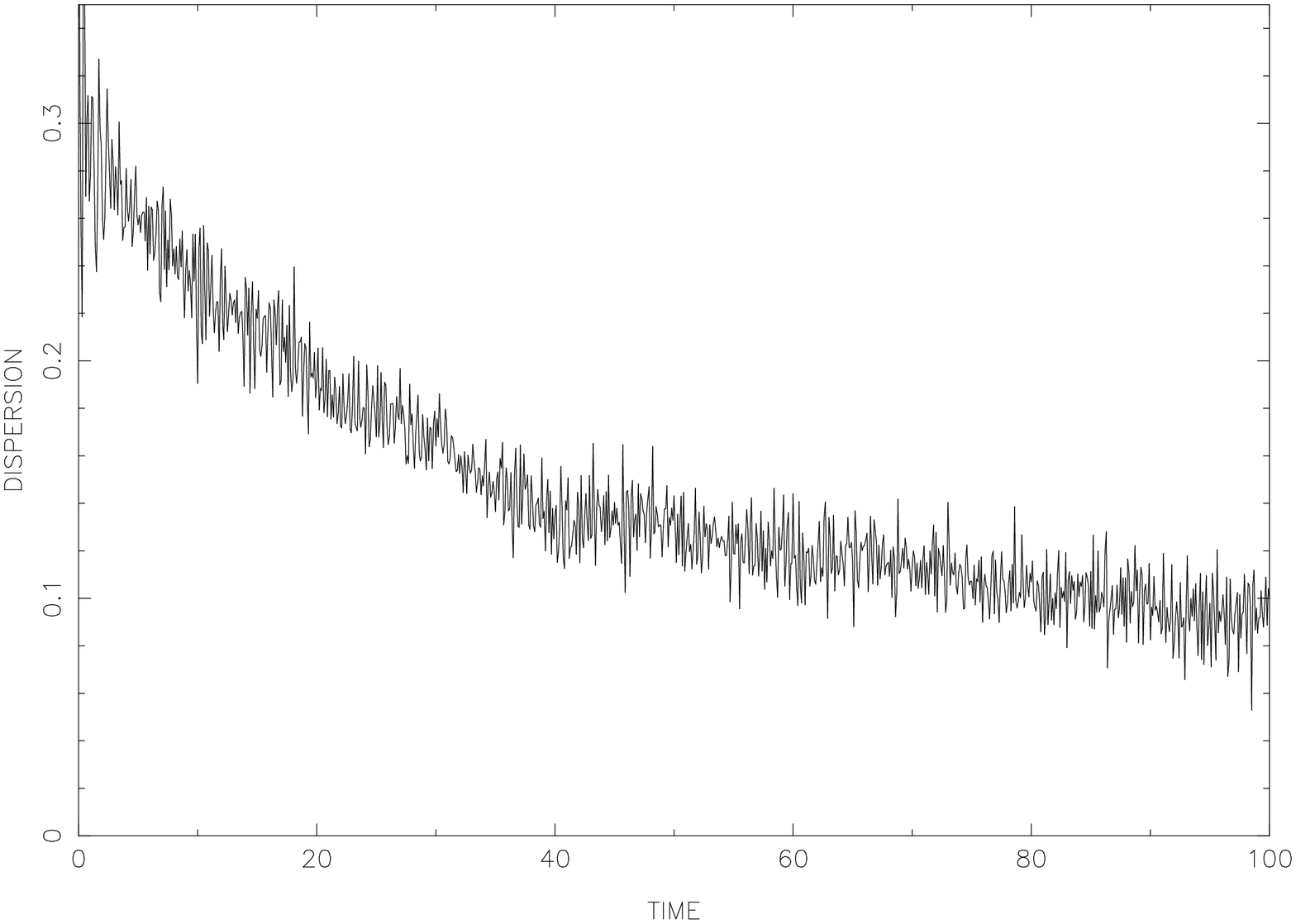,height=5.2cm,width=7.8cm,angle=-90}
\hspace{0.3cm}
\epsfig{file=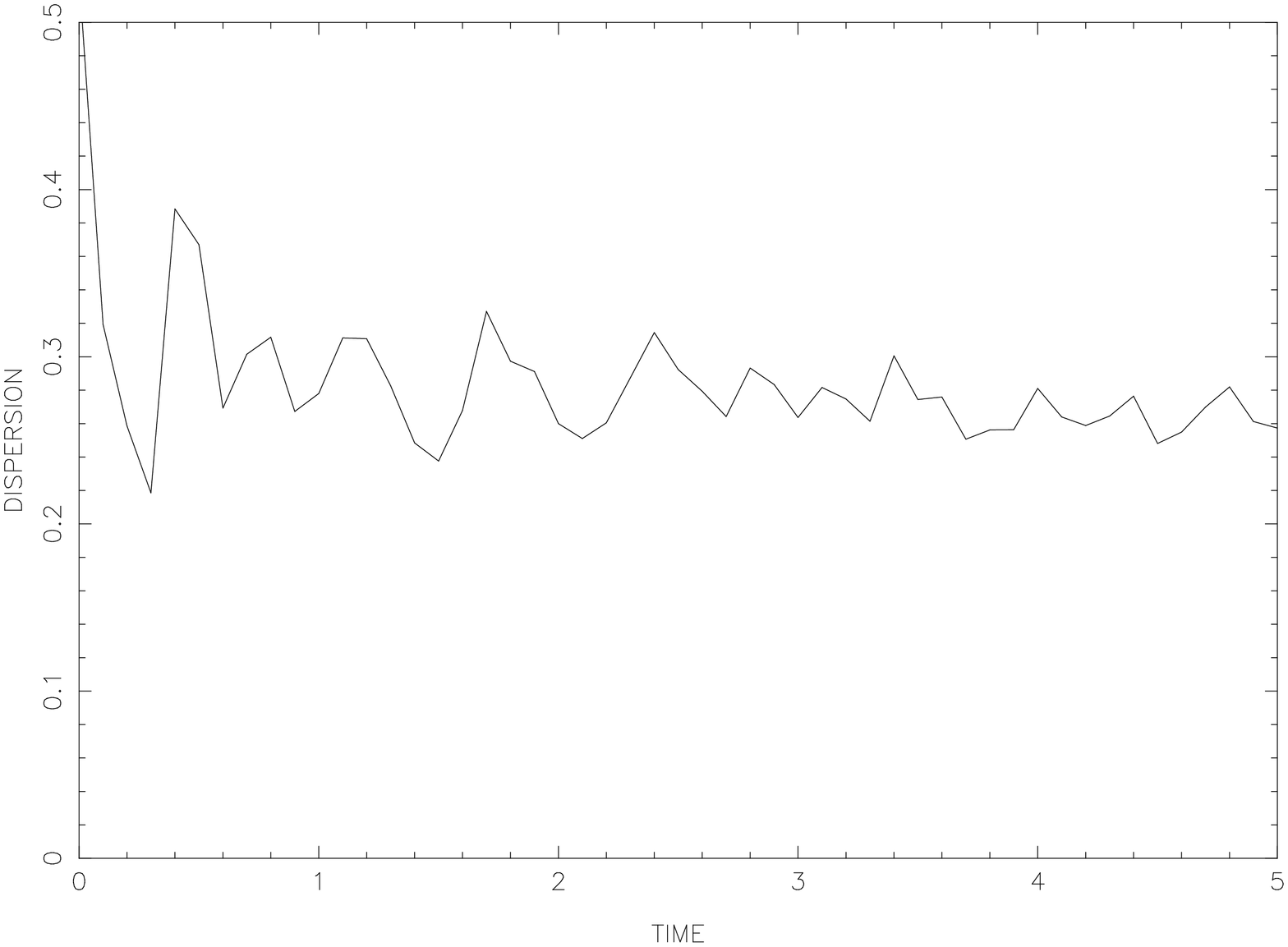,height=5.2cm,width=7.8cm,angle=-90}
\end{center}
\caption{\label{AS1diszm}
Time (in units of $\tau_c$) evolution of the
 relative dispersion of the trace of the velocity dispersion tensor for 
a system 
starting from homogeneous density and isotropic velocity distribution 
decreasing with radius according to Eq.~\protect\ref{sphex:VAI} with $p=1$ }
\end{figure}

Fig.~\ref{AS1diszm} shows the evolution of $\sigma_{d}$ for a system 
starting with
isotropic initial velocities decreasing exponentially with radius 
(according to Eq.~\ref{sphex:VAI}). One can see that $\sigma_{d}$ 
decreases monotonically (give or take random fluctuations) but
only over many crossing times,
and reaches a value of $10 \%$ or so over a time-scale comparable to
the  two body relaxation time of standard relaxation theory. 
According to that  theory,
 the half mass relaxation time is given by~\cite{Spitzer} 
\begin{equation}
t_{rh} =\frac{0.14 \times  2500}{\ln (2500 \gamma_{l})} \times r^{3/2}.
\label{sphex:tbr}
\end{equation}  
After the first few crossing times, during which
 the system moves away from the spatially homogeneous initial 
distribution and 
reaches its slowly evolving dynamical equilibrium state, the 
half mass radius settles down to a value of about $0.65$.  
 Formula~(\ref{sphex:tbr}) therefore predicts a relaxation time-scale
in the range between about $36$  and $45$ crossing times, 
depending on whether one uses the value $\gamma_{l}=0.4$ of 
Farouki and Salpeter~\protect\cite{Farouki and Salpeter2}
 or $\gamma_{l}= 0.11$ of 
Giersz and Heggie ~\protect\cite{Giersz and Heggie}.
Looking at the plot 
in Fig.~\ref{AS1diszm}, one can see that during this time interval 
there is a turnoff 
in the $\sigma_{d}$ time series, so that the decrease in its value with
time is much slower. What happens is that beyond $t=t_{rh}$, 
the velocity gradient
in the central area is, for all practical purposes, smoothed out. 
The remaining
error is due to a slight gradient in the outer regions (which is also 
eventually smoothed out, but on a longer time-scale of about 20 additional
crossing times)~\cite{Note2}.

Thus, it would appear that this system evolves towards the final equilibrium 
on the thermal time-scale, and that this is adequately described by
standard two body relaxation theory. As was mentioned in the introduction,
this theory may be adequate in describing energy relaxation 
(which is presumably
the main mechanism acting here), 
even though the dynamical picture it is based on 
may be rendered inaccurate by the chaotic nature of $N$-body trajectories.
Nevertheless, it has been suggested~\protect\cite{Gurzadyans} 
that the exponential 
divergence of phase space trajectories
accompanying such motion {\em itself} induces a relaxation 
time-scale --- which 
could be different from the standard two-body time since it is obtained 
from entirely different
considerations --- by estimating the time-scale for 
smoothing out (or otherwise radically modifying)
 the phase space density distribution (``mixing''), 
as is suggested
by the ergodic interpretation of statistical 
mechanics~\protect\cite{Krylov,Sagdeev}.
However, the exponentiation time-scale in gravitational $N$-body systems 
is found to be of the order of a dynamical 
time~\protect\cite{Goodman et al.,Kandrup et al.,El-Zant2}, so if it 
did directly correspond to the evolution towards the
thermal equilibrium state,  one would expect such a state to be reached within 
an accuracy of a few percent after a few crossing 
times (see Eq.~29 in ~\protect\cite{El-Zant1}).
The fact that this   is obviously not the case has led some investigators
to conclude that the exponential divergence of $N$-body trajectories 
has nothing
to do with their relaxation~\protect\cite{Goodman et al.} or that  
it is more likely to {\em only} affect the time-scale of 
achieving {\em dynamical}
equilibrium ~\protect\cite{Kandrup3}.
For the system just described,  this corresponds to
 the  time-scale for reaching the slowly evolving state which is
achieved when $\sigma_{d}$ has gone down from an initial
 value of $0.47$ to a value to about $0.3$. 
This happens during the first few crossing times (Fig.~\ref{AS1diszm}). 

The above conundrum may apparently be resolved by noting 
 that, for large-$N$ systems in dynamical
equilibrium, the (mixing)
 time-scale for obtaining an equilibrium phase space density distribution 
is not  simply a few exponentiation time-scales but may instead be given 
by~\protect\cite{El-Zant2}  
\begin{equation}
\tau_{r}= 2 \sqrt{3} \ln (1/d) T \sqrt{N}  \tau_{e},
\label{time}
\end{equation}  
 where $d$ is the  
linear phase space resolution over which averaged (coarse grained)
distribution functions
should not evolve after a time of order $\tau_r$. 
We calculated 
the exponentiation time-scale  $\tau_{e}$ for the systems studied here
 by using the Ricci 
curvature method~\protect\cite{Gurzadyan and Kocharyan,El-Zant1,El-Zant2}.
 It  was found to be about  0.4 crossing times.
The kinetic energy is also $T \sim 0.4$.
Thus, for a resolution of ten  percent say, the 
time-scale given by~(\ref{time}) is $\sim 67$ crossing times, 
which is  compatible with the results described here. 
Thus, it will be necessary to examine the variation of $\tau_r$
with (sufficiently large) $N$ to see if the above formula has 
any validity in determining the evolution towards equilibrium 
or whether the standard theory
holds. The results of such simulations are currently being 
analised~\protect\cite{El-Zant and Goodwin}.

\subsection{The case when the local dynamical equilibrium is close 
to the thermal equilibrium state}

\begin{figure}
\begin{center}
\epsfig{file=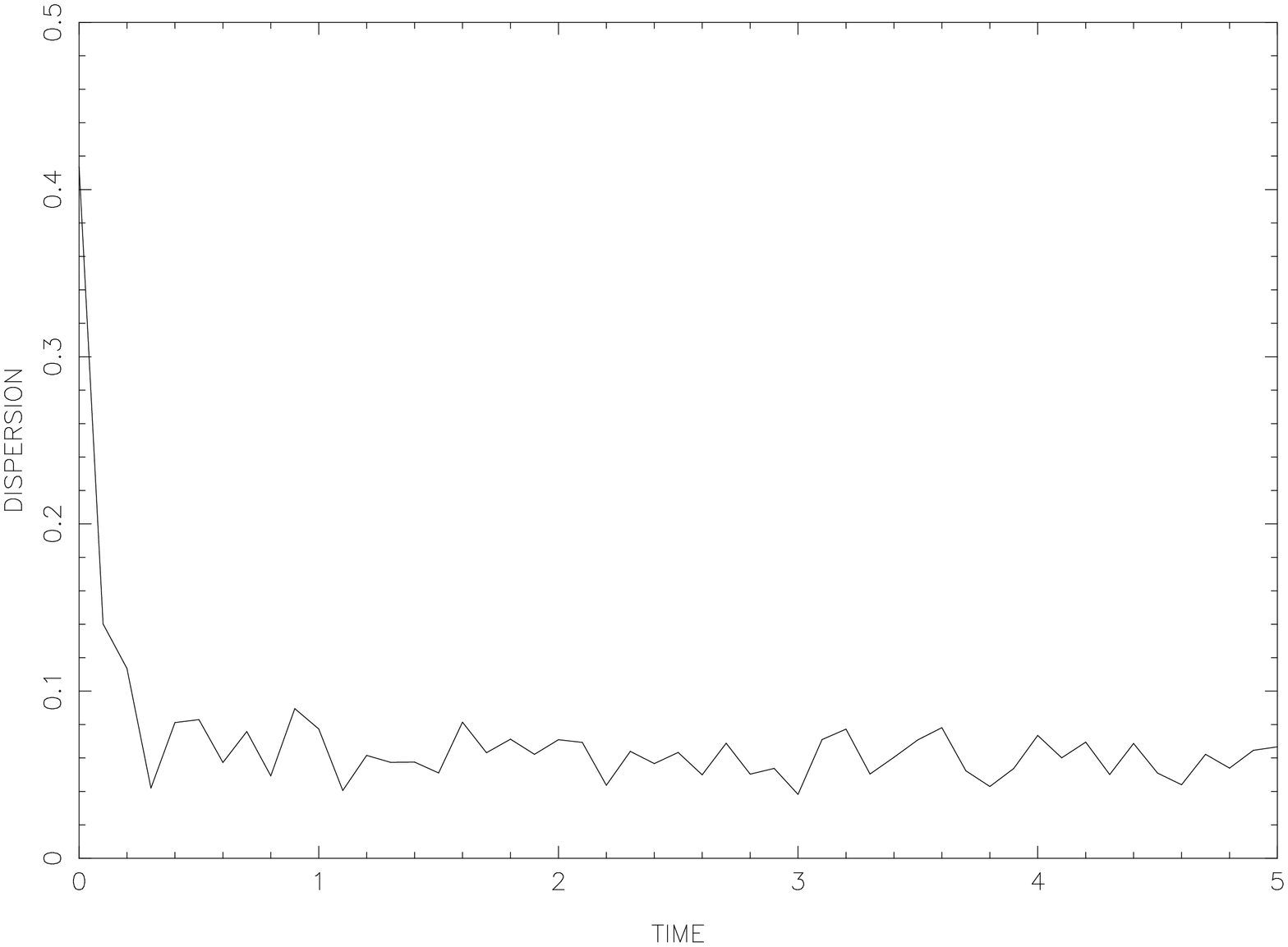,height=5.2cm,width=7.8cm,angle=-90}
\hspace{0.3cm}
\epsfig{file=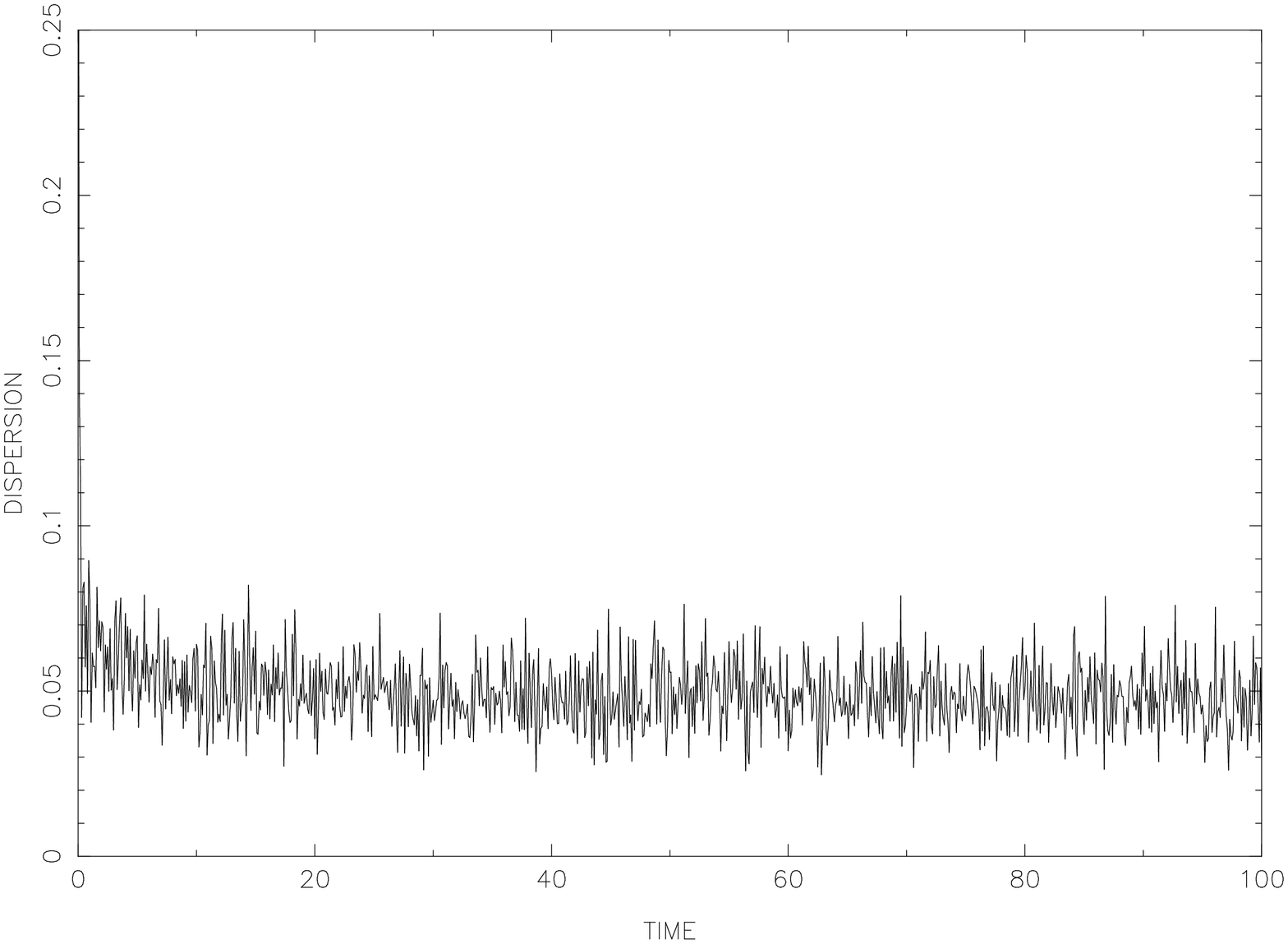,height=5.2cm,width=7.8cm,angle=-90}
\end{center}
\caption{\label{AS2diszm}
Time (in units of $\tau_c$) evolution of the
 relative dispersion of the trace of the velocity dispersion tensor 
for a system 
starting from homogeneous density and isotropic velocity distribution 
increasing with radius according to Eq.~\protect\ref{sphex:VAI} with $p=-1$ }
\end{figure}

The situation was found to be very different for the system starting 
with a temperature 
inversion (i.e., with $p=-1$ in Eq.~\ref{sphex:VAI}). 
In this case, the long lived steady state  arrived at on a time-scale
comparable to the exponentiation time-scale {\em is} the thermal equilibrium
state. As can be seen from Fig.~\ref{AS2diszm}, this state is reached in 
essentially less
than a crossing time. After that $\sigma_d$ deviates from zero by only a few 
percent. This simply corresponds to the particle noise.
And it  is to be stressed  that here, like in the case
when the initial velocities decreased with radius, the virial ratio remained
 nearly constant throughout the evolution. 
The density distribution and
parameters that sensitively depend on it (like the potential energy)
 changed only slowly, with a rate comparable to that of the
slowly evolving system described in the previous subsection. 
We conclude therefore that this 
equilibrium state was
not reached through the conditions normally associated with violent 
relaxation~\protect\cite{Saslaw}. It also could not have been reached through 
standard two body relaxation, since  this process takes place on  the
much longer time-scale of Fig.~\ref{AS1diszm} . 

It has been argued~\protect\cite{El-Zant2} that the exponential 
divergence of trajectories
of gravitational systems
in dynamical equilibrium mainly serves to smooth out the ($6 N$)
 phase space density 
distribution
on subspaces compatible with the dynamical equilibrium and cause only slow 
diffusion
away from that state.
Thus while Eq.~(\ref{time}) gives the relaxation time-scale for
 a system where 
the phase space 
divergence of trajectories is in such a way as to maintain a given dynamical 
equilibrium 
 --- i.e. by covering different configurations of a slowly evolving 
equilibrium ---
if such an equilibrium is not found, the divergence can take place at 
arbitrary directions
in the phase space. 
 Because of the exponential divergence normal 
to the phase space motion, 
 a dynamical equilibrium 
with smooth macroscopic density and velocity distribution may be quickly 
attained  even if the conditions for violent relaxation 
(which require large density
fluctuations and far from equilibrium evolution) are not satisfied.
 Once a dynamical equilibrium is found, the geometry of the configuration 
space must
behave in such a way that {\em globally} the exponential divergence is 
effectively
{\em along} the motion (although locally it must be normal). 
 Only the small scale fluctuations away from the equilibrium
 due to the discreteness noise 
modify this. 
The evolutionary time-scale related to these is much larger 
however~\protect\cite{comamr}.

It is worth noting here that behaviour similar to what has been observed here 
occurs for 1-dimensional gravitational systems. In this case too, systems
 prepared near stable dynamic equilibria gradually approach the thermal 
equilibrium distribution due to collisional 
relaxation~\cite{Tsuchiya et al.,normal}, while 
those initialized from other  configurations rapidly 
relax to a state very close to thermal equilibrium~\cite{inverted}.

\subsection{The invariance of the final thermal equilibrium state}

The final equilibrium state should be, by definition,
 invariant under the effect of gravitational interactions between
particles, 
and therefore, if it is thermodynamically stable, one 
does not expect it to 
evolve~\cite{Lynden-Bell}.
As can be seen from Fig.~\ref{AS2diszm}, this is actually the case. 
After the initial
relaxation phase,
 during which $\sigma_{d}$ drops from its initial value (of about $0.32$)
to an average value of about $5 \%$  (consistent with fluctuations 
arising from the particle noise), the
value of $\sigma_{d}$ hardly evolves at all.

\begin{figure}
\begin{center}
\epsfig{file=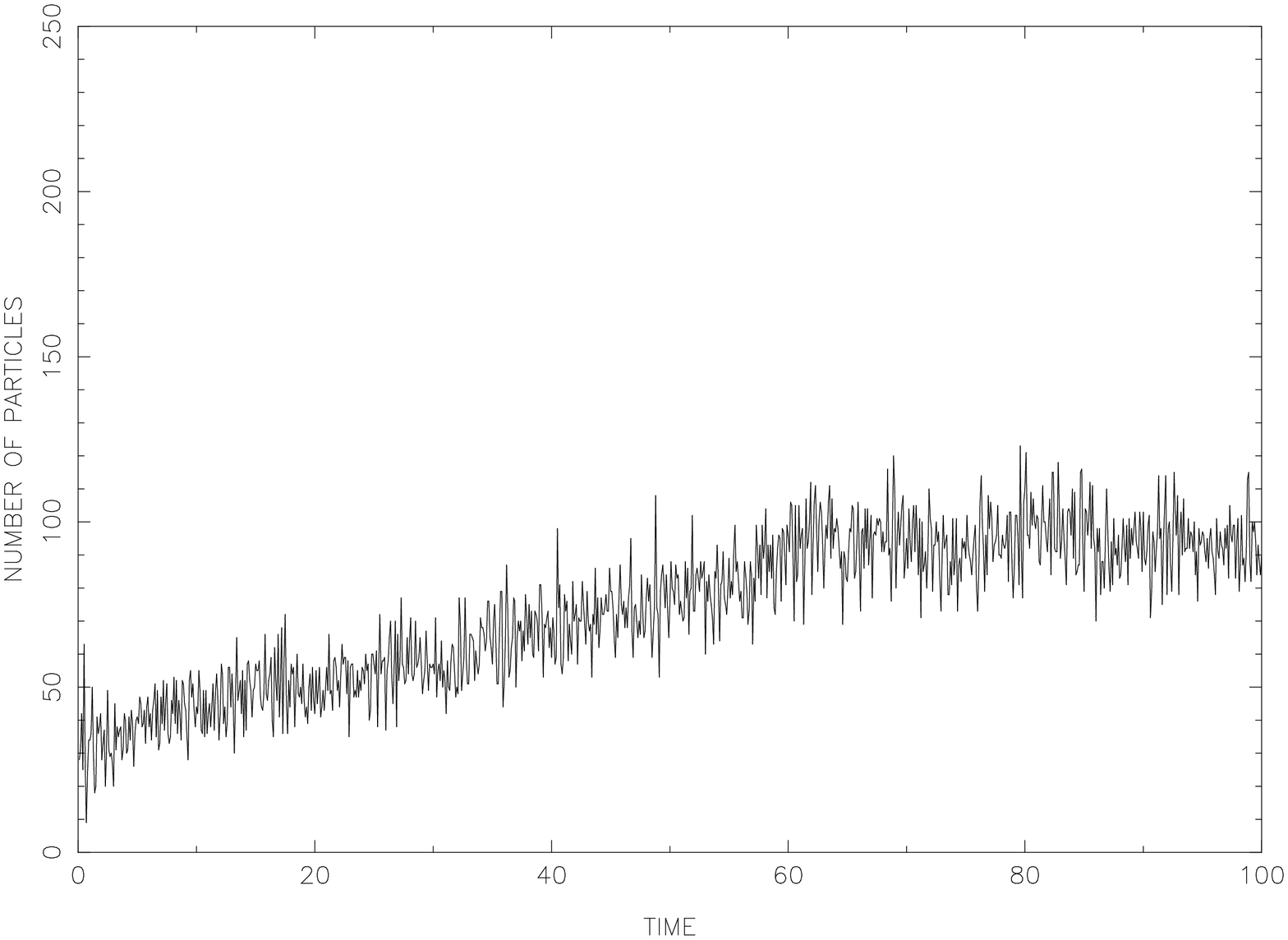,height=5.2cm,width=7.8cm,angle=-90}
\hspace{0.3cm}
\epsfig{file=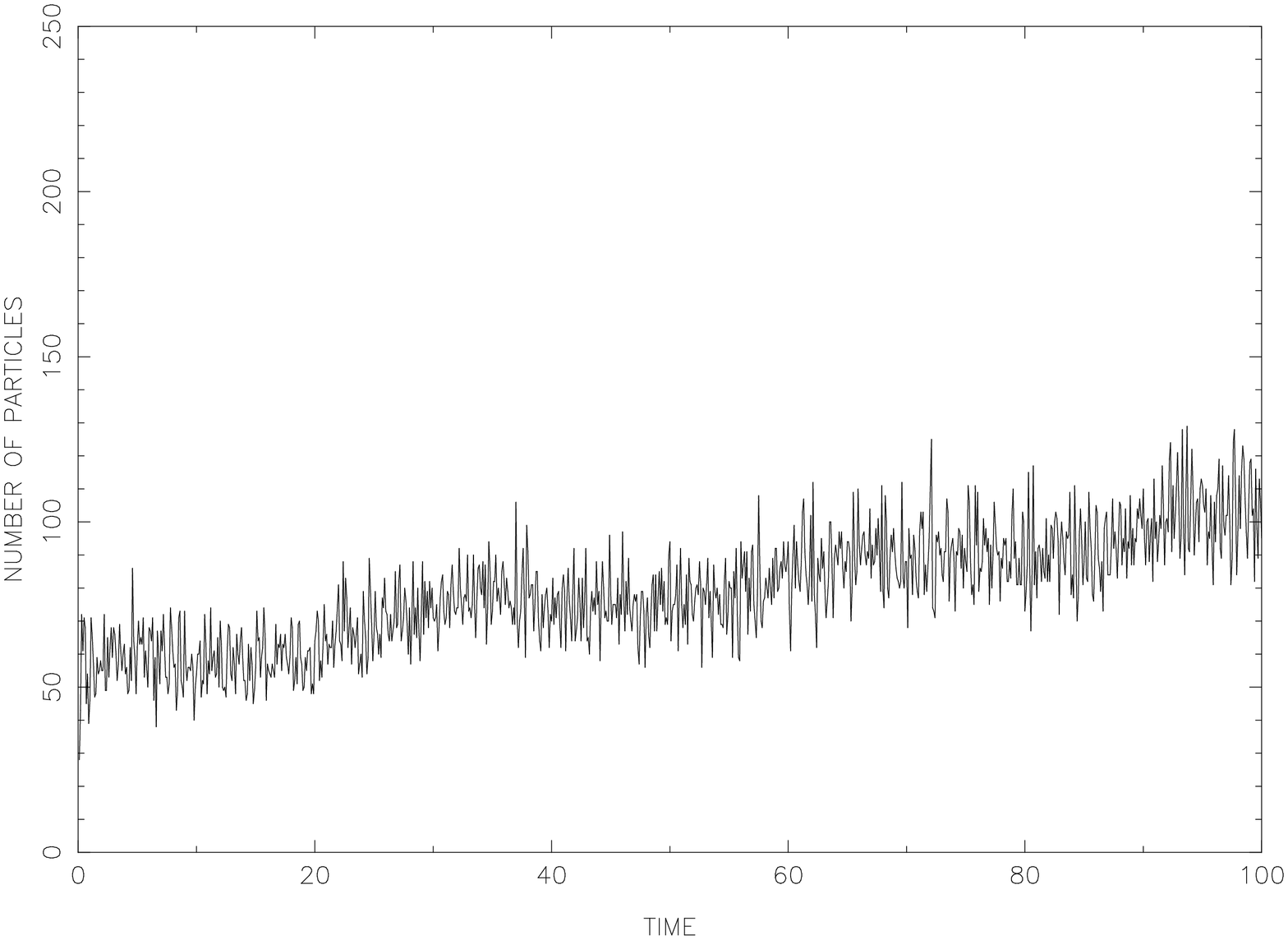,height=5.2cm,width=7.8cm,angle=-90}
\end{center}
\caption{\label{AS1CNS}
 The evolution of the number  of particles in the 
central cell ($r=0.2 \times r_{0}$) as a funtion of the (crossing)
time, in  runs corresponding to the 
plots in Fig.~\protect\ref{AS1diszm}   
(left) and Fig.~\protect\ref{AS2diszm} }
\end{figure}

Further evidence that this final state is long lived and that no
 gravothermal
catastrophe occurs can be obtained by looking at the evolution of the 
particle number 
in the central spatial cells of the two systems we have considered. 
This is shown in 
Fig.~\ref{AS1CNS},
after an initial increase as the system settles towards the isothermal 
configuration 
(starting
from the homogeneous density distribution), the particle number in the 
central cell
 evolves only slowly, 
and then only in a limited way which does not signal any 
gravothermal 
collapse, but instead an evolution towards a more detailed equilibrium. 
Thus while evolution towards the equilibrium distribution in velocity
space is fast, evolution in configuration space is relatively slow.
This again illustrates the multiple time-scales present in gravitational 
systems.
For the  system starting from the initial state where the 
velocities decreased outwards, the central density ceases to evolve 
significantly after  about sixty crossing times.

For every set of numbers $M$, $E$ and $R$,
there exist a unique stable   
 Lane-Embden isothermal sphere configuration(cf Fig. 8-1 of Ref.~\cite{BT}).
 In our runsO
  $M=1$ and   $E=-0.185$.
Since very few particles (usually not more than 10) are present beyond 
$r-r_0=0.2$, we take this to be the boundary value $R=1.2$.  
In this case, the central density of this unique isothermal sphere is 
given by~\cite{BT}
\begin{equation}
\rho_0 = \frac{3 \sigma} {4 \pi G r_{c}^{2}}, 
\end{equation}
where, in the units used in this paper, $G=1$. For the above value of the
total energy and $vir=1.4$, the velocity dispersion $\sigma=0.186$.
Here $r_c$ is the 
core radius of the system (within which the density is roughly constant). 
For the above values of the parameters 
$r_c \sim 0.3 R$
(this, in fact,  can be inferred from Table 4-1 and Fig. 8-1 of 
Ref.~\cite{BT}). One can then  deduce that inside 
$r \leq 0.2$ there will exist about a hundred particles of mass
$1/2500$ . This is indeed approximately the number of particles found in 
the central cell (that is for $r \leq 0.2$) at the end of our simulations, 
which suggests that the resulting end states
 are indeed  thermal equilibria.

\subsection{The approach to equilibrium in anisotropic systems}

To conclude this section, we finally note that
  the evolution of $\sigma_{d}$ for the system starting
with anisotropic velocities (prescribed according to Eq.~(\ref{sphex:vani})) 
turns out
to be intermediate between the two cases described above. The dispersion 
decreases within a couple of dynamical times to a value of about $12 \%$, 
after that the system continues to evolve to a more precise thermal
equilibrium  state
 but in a 
less rapid manner, taking about 65 crossing times  for the dispersion to hover
around an average $\sigma_{d} \sim 5 \%$.
The initial anisotropy is however washed away within
a time-scale of $10$ to $20$ crossing times. This a prelude of things
 to come (Section~\ref{anis})
when it will be confirmed that the relaxation of velocity anisotropies 
appears to be
much faster than that of the energies.

\section{Relaxation of  velocity anisotropies}
\label{anis}

Energy relaxation is one of the effects produced by discreteness noise 
in $N$-body 
systems, but there are others. An important one 
is the relaxation of initially
anisotropic velocities. This, in a spherical system, will correspond to 
angular momentum relaxation.
As was mentioned in the introduction, it is not evident that the energy 
relaxation
time-scale, which should correspond to the long relaxation
 time-scale of Section~\ref{slowap}, 
is the only time-scale  of interest when examining  effects induced 
by the discreteness of $N$-body systems. It is important therefore to check
how other quantities, which may be more directly related to the detailed 
particle trajectories, tend towards equilibrium values.

It was seen near the end 
of the previous section that the relaxation of 
velocity anisotropies  can be   
much faster than the thermal (energy relaxation) time-scale
 when the anisotropy is initially mild.
In this section we study the evolution of velocity dispersions for systems 
where these are initially strongly anisotropic. To do this, we simply start 
our system 
 with the $y$ and
$z$ velocities set to zero. 
The $x$ velocities, which do not vary with radius,
are scaled in such a way as to keep the same  virial ratio 
  as before. The system starts from homogeneous density and 
will therefore have the same energy as the ones discussed in 
previous sections.
 This configuration  
is obviously very far
from dynamical equilibrium and our system ``violently relaxes'', 
with the viral ratio
oscillating wildly for many dynamical times. 
It settles to a value of about 0.73 ($vir=1.46$) after 
roughly ten crossing times, this will therefore be our zero point 
in calculating the relaxation time. 

After the ten initial crossing times, the trace of the velocity dispersion
 tensor does not 
vary significantly with radius 
($\sigma_{d} \sim 6 \%$ almost independent of $r$), 
and in this sense 
the system is isothermal. Moreover, this is mostly true in each direction 
$x$, $y$ and $z$
{\em independently}. We can therefore average our anisotropy measure over all
radii to get less noisy data. For easy comparison with energy relaxation, 
we will
use for this purpose 
the same formula~(Eq.~\ref{disperse}) as  before, replacing the trace of the 
dispersion tensor in each radial
cell ($\sigma_{t}^{l}$)
 with the sum of the velocity dispersions over all cells in the $x$, $y$ and
$z$ directions. The number 10 is then replaced by 3 and the $\bar{\sigma}$ 
is taken as 
the average of $\sigma_{x}$, $\sigma_{y}$ and  $\sigma_{z}$.

\begin{figure}
\begin{center}
\epsfig{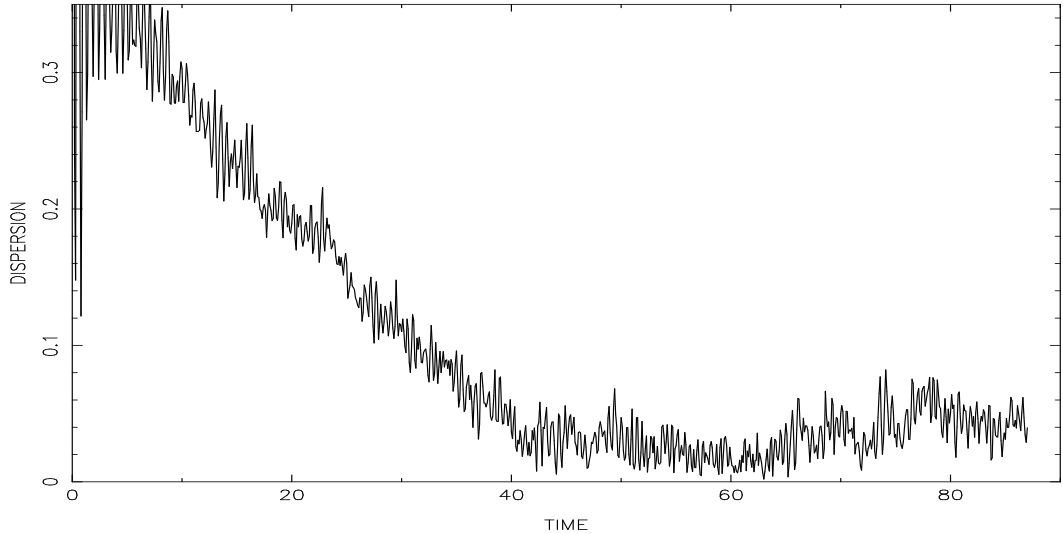}
\end{center}
\caption{\label{anisotropic}
Relaxation of initial velocity anisotropies. 
The system is started with zero $y$ and
$z$ velocities and achieves a quasisteady state after $\sim 10 \tau_{c}$}.
Time is measured in units of the crossing time $\tau_c$
\end{figure}

The results are shown in Fig.~\ref{anisotropic}. After the initial 
``violent evolution'' of the
first ten crossing times, following which the dispersion settles to a value 
of $\sim 0.32$,
it starts decreasing abruptly (almost linearly!), reaching the equilibrium
(isotropic) value
after a total of $\sim 40$ crossing times. This means that the {\em effective}
relaxation time (after achieving dynamical equilibrium) towards isotropic
velocities is about 30 crossing times. This is significantly 
smaller  than the  time required for complete energy relaxation 
(about a quarter of that time). Thus it would appear that, at least for the 
case of  closed systems  (and for the initial conditions examined here and 
in Section~\ref{slowap}),
 relaxation of initial anisotropies 
 seem to be {\em much faster} than energy relaxation.

Three obvious and related questions will however have to be addressed 
before the above conclusion
is credible. First i) does the system become triaxial
during its evolution, and, if so, can much of the relaxation be due an 
abundance of chaotic
orbits in the mean field? ii) if the system is not spherical during its 
evolution, what is the effect 
of the spherical enclosure on the relaxation of such a system?
iii) is the density distribution in the system studied in this
section and the ones for which energy relaxation was examined
 similar? (this is important since the 
relaxation is expected to depend on the local density). 

To answer the first question,
we have calculated the potential energy tensor~\protect\cite{BT} during 
the evolution. From
this we infer the axis ratios (assuming an ellipsoidal density distribution
with constant axis ratio). 
The answer is yes, as may be expected, 
a system starting
from such initial conditions is not spherical during its early evolution. 
However, it is
only mildly triaxial at 10 crossing times (with the two longest density
 axes having ratios $a/b=0.88$, which means  still larger equipotential axis 
ratios)
and is almost completely axisymmetric ($a/b=0.96$) after 20 crossing times. 
Since such mild 
triaxialities are not likely to cause rapid evolution 
(without strong central mass 
concentration~\protect\cite{Merrit and Quinlan}), 
and since the evolution does not  slow
down after the asymmetry
is almost completely lost (at $\sim 30$ crossing times), and the 
departure from sphericity
is very small, we conclude that such mean field evolution would not 
appear to play
a central part in relaxation process.

One may suspect that the presence of an artificial spherical box 
surrounding an asymmetric system
may play a role in its evolution towards a spherical shape, 
which may be accompanied by velocity
relaxation towards isotropy. However  this does
 not appear to be the 
case here: since the whole process of evolution towards 
isotropy takes 
place on a fraction of an energy relaxation time, one expects most particles 
to conserve their 
energy and be confined  away from the spherical enclosure 
(by their zero velocity surfaces). 
In this case, particles in the outer
areas would be the most affected and would then ``communicate'' the 
disturbance felt to the inner areas. It follows that one would  expect the 
evolution towards
isotropy to take place from outside in --- i.e., the outer regions become 
more isotropic 
before the inner ones. This was however not found to be the case, 
if anything it was the inner areas 
which evolved slightly faster (presumably due to their higher densities).

In principle, by virtue of the uniqueness of isothermal spheres for a
 given value of the 
parameter $\mu$, the density distribution of all final equilibrium 
configurations
with the same total energy, radius and mass should be the same.
However, we have allowed for our 
boundary to be elastic (thus $R$ is not completely fixed) and the energy 
in our $N$-body
simulations is not exactly conserved. Therefore, the density profiles may 
vary slightly from
one model to another. In particular, one expects the system studied in this 
section to be more
centrally concentrated due to its initial violent 
relaxation~\protect\cite{Soker}. 
 This indeed turns out to be the case,
 the present system  is more centrally concentrated. 
However it is only significantly so in the inner cell or two.
 Beyond the fifth
cell the reverse is actually true.
 We can again appeal to the uniformity
of the relaxation towards isotropy at all radii (including the mid regions 
where the density 
in both systems is similar) to suggest that the increased central density 
is not an important effect.

In addition, all the above problems do not occur for the system with mild 
initial velocity anisotropies and gradients (discussed at the end of 
Section~\ref{slowap}). 
This configuration stays
almost spherical during the evolution, thus the relaxation towards isotropy 
means relaxation
in angular momentum (which is conserved by our boundary force law).
 This takes place 
{\em concurrently}
with the relaxation of a thermal gradient (so that there is no question of 
different density
distributions affecting the relaxation rate).
Thus, it appears that the effect of relaxation towards isotropy being 
faster than energy relaxation
is real. However, it is still left to future more precise simulations with 
more controlled conditions,
and exploration of the parameter spaces and variation of  particle numbers 
to confirm this potentially important result. 
 
The idea that the relaxation of the velocities towards a Maxwellian 
(which is what
happened here when an isotropic state was reached) may be faster than 
the process 
of energy relaxation is actually quite 
old~\protect\cite{Prigogine and Severne,Kandrup4}. 
In their study, Prigogine and Severne 
used a kinetic formulation that avoided the artificial cutoff introduced 
to eliminate the  
divergence in the Coulomb logarithm. They find that the relaxation towards
 a Maxwellian
is  $\sim \ln N$ times faster than energy relaxation. 
This is compatible 
with what is found here. Moreover, their analysis also suggests the 
oscillations 
around the equilibrium suggested by the longer time evolution of the 
time-series in Fig~\ref{anisotropic}.

\section{Evolution of multimass systems}
\label{multi}

The main difference between a single mass system and one consisting 
of particles
with different masses is that, in the latter case, equipartition of 
energy will  cause
heavier mass particles to spiral towards the centre of the system and reside 
there.
General thermodynamical considerations show that once the total mass, energy 
and radius
of an isothermal sphere are fixed, there corresponds only one configuration 
which
is stable, and this has a unique value for the ratio of the density at the 
centre
and at the boundary (cf. Ref. \cite{BT} Fig. 8-1). 
One therefore expects that a perturbation 
which 
conserves $\mu$ (cf. Eq.~(\ref{sphex:gravo})), but changes the ratio 
of central 
to boundary densities, 
will necessarily cause a system to
move away from this unique thermal equilibrium configuration. And if the 
system cannot move back (by restoring the initial density ratio) the 
equilibrium is
unstable. This is of course precisely the situation here, 
where there is a mechanism that acts as to cause high mass particles to
spiral to the centre leaving $\mu$  unchanged. 
Since this process is due to equipartition 
of energy between particles, it is  statistically irreversible.
Gurzadyan et al.~\protect\cite{Gurzadyan et al.} have  examined 
the evolution of the transition from the case when a thermal
equilibrium   existed and the case when it did not exist. 
The bifurcation
between these cases was found to depend on the existence of a central mass.
 In particular, it was found that as one increased the
central (point) mass, the value of $\mu$ 
that is required to achieve a stable isothermal thermodynamic
equilibrium became significantly larger. That is, the aforementioned curve 
in Ref.~\protect\cite{BT}
is shifted upwards. This means that, other parameters staying constant, 
one has to 
continually increase the system's energy  in order to maintain thermal 
equilibrium as the central density is increased. 

\begin{figure}
\begin{center}
\epsfig{file=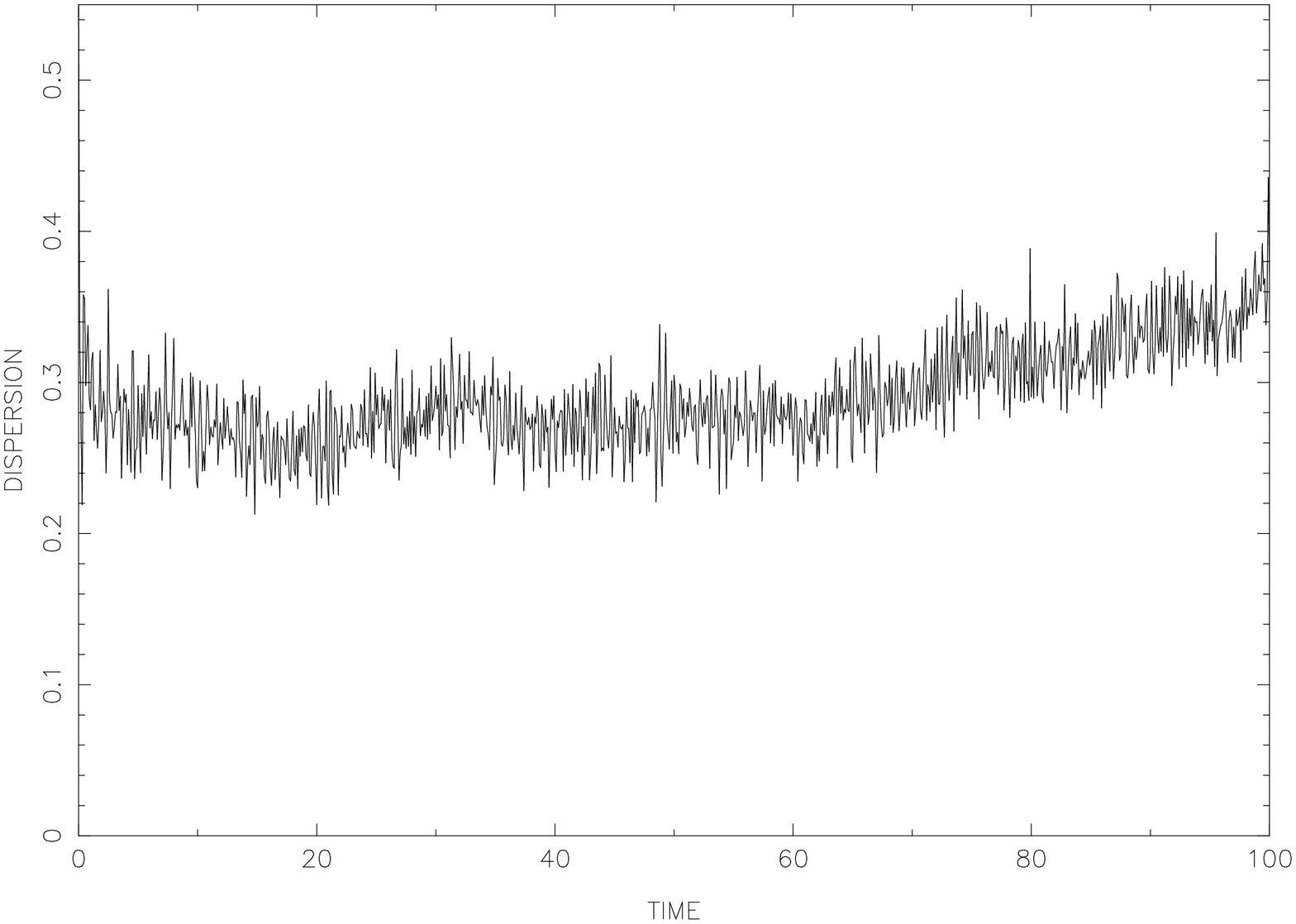,height=5.2cm,width=7.8cm,angle=-90}
\hspace{0.3cm}
\epsfig{file=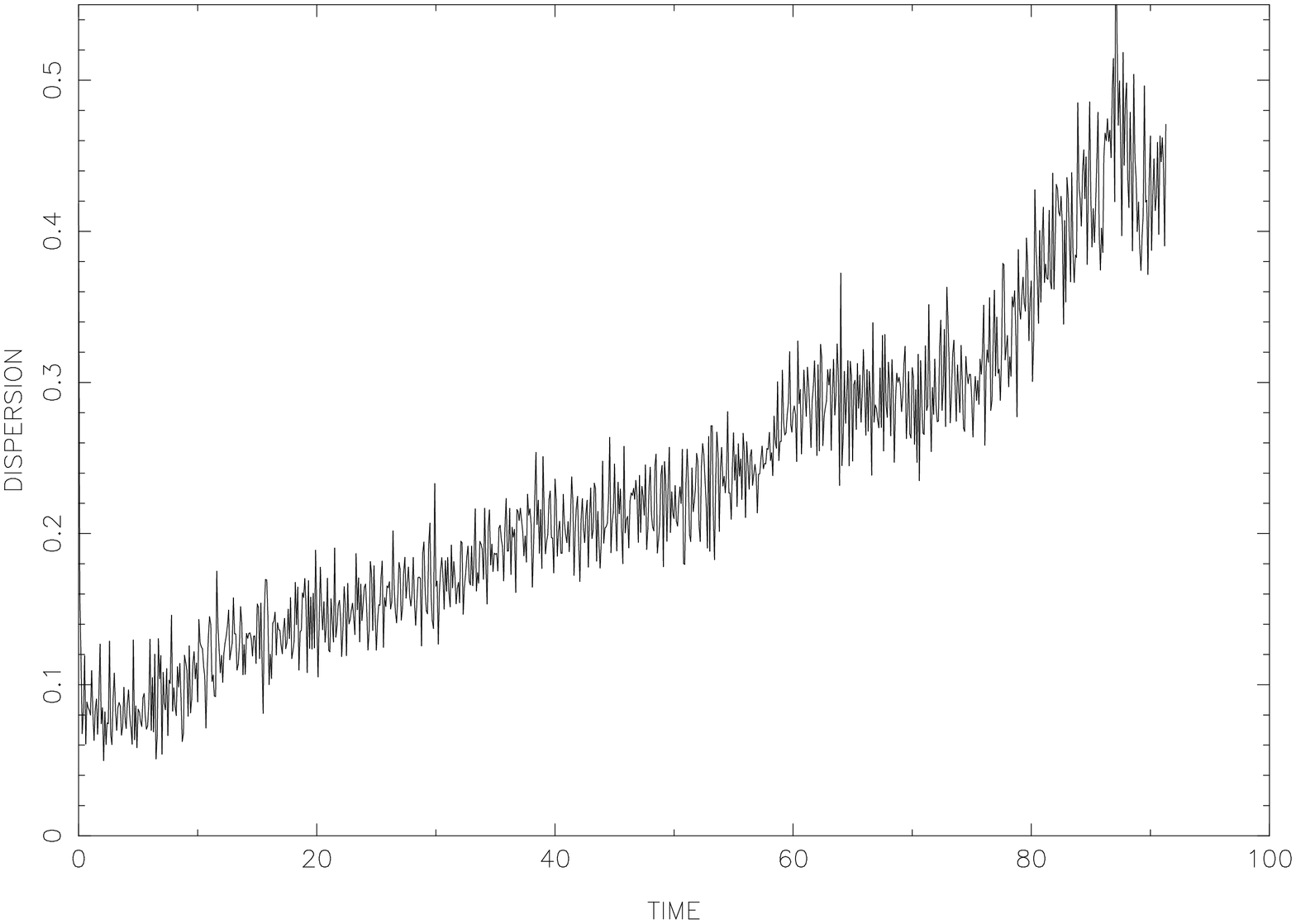,height=5.2cm,width=7.8cm,angle=-90}
\end{center}
\caption{\label{AS1DMdis}
 Evolution of the relative dispersion of the trace of the kinetic energy 
tensor of the   multimass systems of corresponding to 
Fig.~\protect\ref{AS1diszm}
(left) and Fig.~\protect\ref{AS2diszm} as a funtion of the crossing time . 
Left: system started with velocities
decreasing with radius. Right: system started with temperature inversion}.
\end{figure}

Thus, one expects the 
increase in the central density  with
time during the evolution of isolated multimass systems to lead  these 
in the direction of  gravothermal collapse.    
 Fig.~\ref{AS1DMdis} 
shows $\sigma_d$  
for the  models
with velocities scaled according to Eq.~(\ref{sphex:VAI}),
and where the mass distributions  follow 
 a Salpeter  mass function~\protect\cite{Gilmore et al.}   
with the highest mass particle being 
ten times more massive than the lightest particle. 
As can be seen, neither of the models
does tend towards an thermal equilibrium.

\begin{figure}
\begin{center}
\epsfig{file=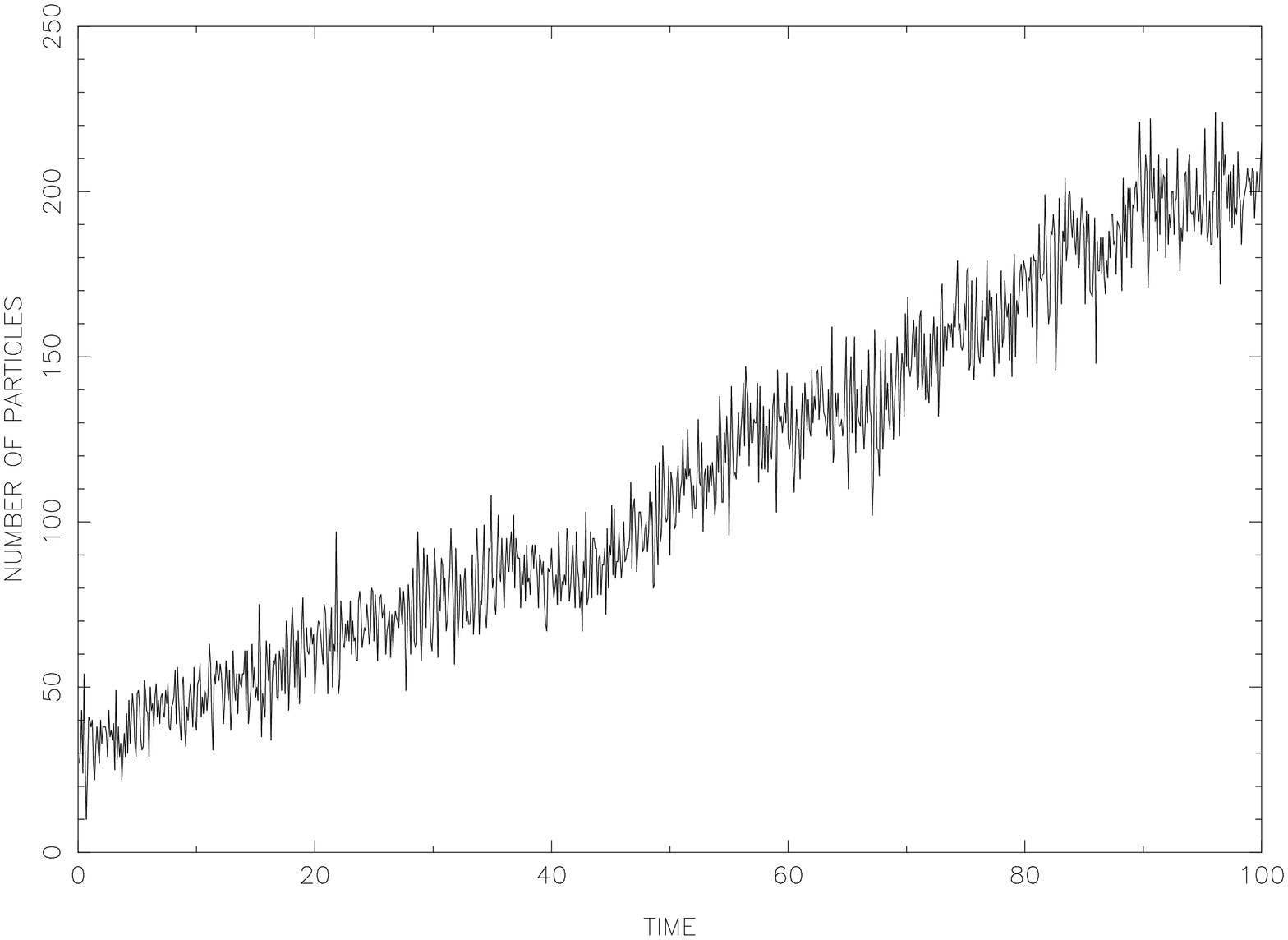,height=5.2cm,width=7.8cm,angle=-90}
\hspace{0.3cm}
\epsfig{file=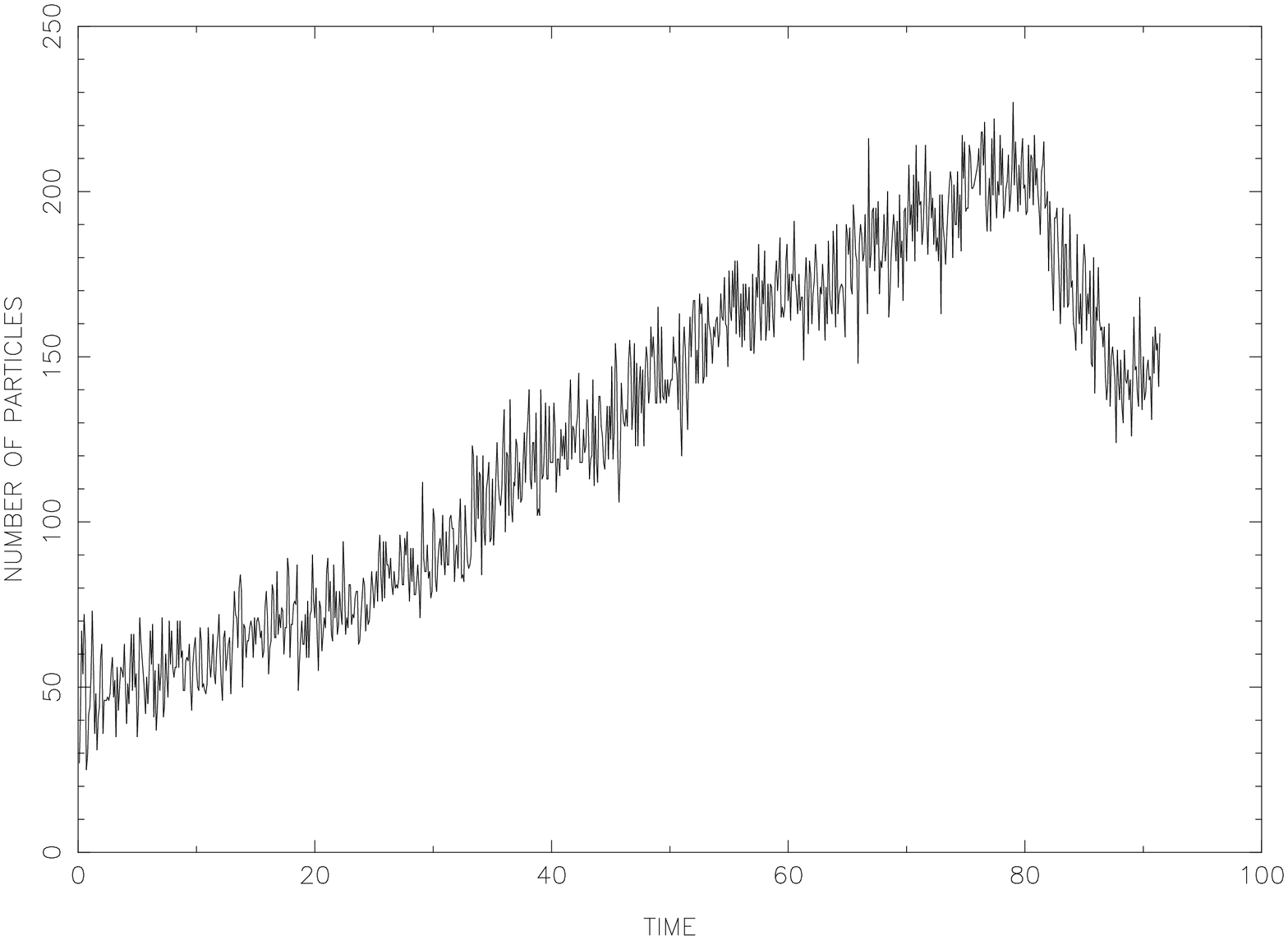,height=5.2cm,width=7.8cm,angle=-90}
\end{center}
\caption{\label{AS1CN} Evolution of the number of particles in the 
central cell
 ofthe multimass systems of Fig.~\protect\ref{AS1DMdis} in terms of the
crossing time. 
Left: system started with velocities
decreasing with radius. 
Right: system started with temperature inversion }
\end{figure}

It is found that, by the end of the run the average mass per particle inside the 
central cell is about twice the original value (corresponding to a sample 
with random mass distribution).
Also, due to the core contraction, not
only is the average mass per particle in the central region 
increasing with time, but also the total number of
particles of any mass (compare Fig.~\ref{AS1CN} with Fig.~\ref{AS1CNS}). 
This process  was found to be  much faster in  the system with initial
temperature inversion --- which tended 
fastest towards  equilibrium in the equal mass case. In fact, for that
system, the number of central particles increases to the point where a large
 number
of tight binaries form, changing the effective specific heat of the system 
and causing
the central density to eventually decrease (the core re-expands). 
The tightening of these binaries causes numerical difficulties and  
the simulation had to be stopped (since the NBODY2 
code is not designed to deal with such situations~\protect\cite{Aarseth}).
The main reason the evolution is faster for the case when 
thermal equilibrium 
is rapidly reached 
is probably  because this equilibrium is more centrally condensed
than the  intermediate dynamical equilibrium states the thermally evolving 
system passes 
through. The higher density means that the accumulation of heavy mass 
particles 
happens at faster rate, which in turn increases the density and velocity 
gradients and so on...

It is important to note here that, although the concentration of the heavier
mass particles towards the centre is due to those loosing kinetic energy in 
gravitational encounters with lower mass ones, true energy equipartition is 
never achieved. This is because a stationary thermal equilibrium is never reached.
Instead, because the system suffers a gravothermal catastrophe, it continually 
evolves towards more inhomogeneous distributions  
(in velocity and configuration spaces) with the kinetic energy of particles 
continually varying. The total kinetic energy increases as the core contracts and
the virial ratio increases, 
if anything the variations in the kinetic energies of the particles are enhanced.
This is in contrast to the 1-dimensional case  (which don't display
unstable  thermal 
properties and) where equipartition can be achieved~\cite{yawnlett}.

The above results suggest that there are, in principle,
statistically  irreversible effects
 that can change the density distribution of a given system away 
from that 
of the stable isothermal solution compatible with its 
radius, total energy and mass. 
This in turn suggests that 
condition~(\ref{sphex:gravo}) is only a necessary condition for a system to
evolve to thermal equilibrium. 
It has been suggested~\cite{Kiessling} that the aforementioned
condition guarantees only that an isothermal sphere is a {\em local}
free energy minimum. It then simply happens that these 
more concentrated states are dynamically inaccessible 
(or at least have low probability of being reached) 
from certain initial configurations. 
This is obviously the case 
for the equal mass systems discussed in the previous section. 
There could be however different configuration of equal mass systems
which do not evolve towards the thermal equilibrium  state even when it is predicted 
to exist. This assertion will be tested and discussed further 
in Section~\ref{phatran}.

\begin{figure}
\begin{center}
\epsfig{file=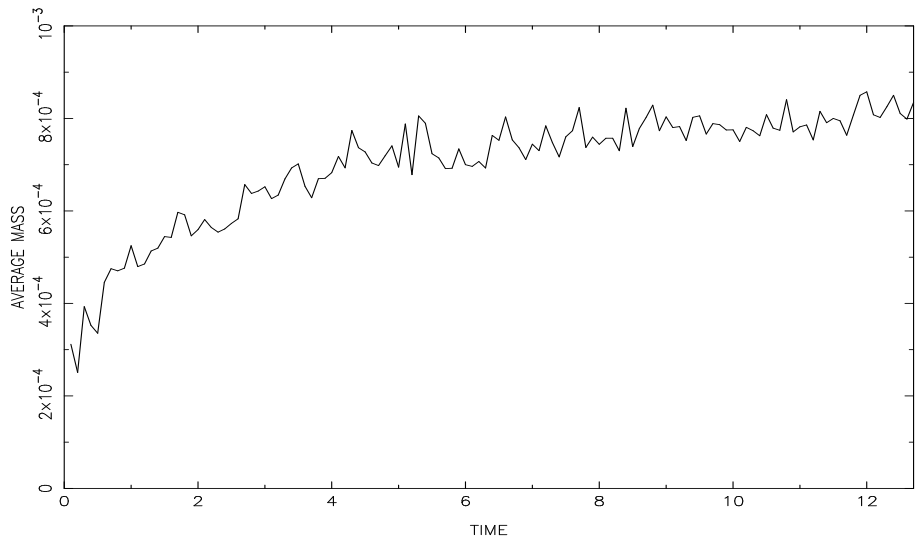,height=7.0cm,width=12cm,angle=-90}
\end{center}
\caption{\label{mulmass}
Average mass per particle inside $0.1 r_{0}$ for a multimass
 system where the virial ratio varies significantly
during the evolution. The total mass of the systems is unity.
Time is measured in units of the crossing time $\tau_c$}
\end{figure}

It was observed
that the process of increasing the average mass per particle  and the 
 total mass
in the central area is effective over a time-scale much smaller 
than the standard two body relaxation time. 
This was especially so for systems starting from initial conditions that, 
although in virial equilibrium,
are far from dynamical equilibrium, so that significant change in the virial 
ratio occurs during the subsequent evolution. 
We ran a simulation, for example, in which the initial velocity 
distribution was given by Eq.~(\ref{sphex:vani}) 
(i.e., anisotropic velocities with asymmetry increasing
towards outside) but with the exponential factor $s=4$. 
During the first two  crossing times or so, $vir=2T/V$, which starts 
from the
usual value of 1.38, fluctuates between values of   1.24  and  1.68. 
During this time
the particle number in the central cell ($r < 0.2 r_0$)
rises from an initial value of 27 to 
about 400.
After that   $vir$ settles down to a value of about 1.54 which 
changes 
relatively slowly (to about 1.6 after $\sim 13$ crossing times) and 
the number 
of central
particles settles down to about $250$.
Fig.~\ref{mulmass} 
shows the average particle mass inside the sphere with radius equal to half
of that of the central spatial cell ($r < 0.1 r_0$) which is clearly
seen to increase over a few crossing time. 
The  mass per particle was found to be   above
average in all the inner cells and below average beyond the fifth cell.
In these outer cells the average mass is $\sim 0.35$ after a few dynamical
times. We stress here that this was found to be
 true even for the very outer cells where the
the standard relaxation rate is relatively long.

One possible explanation 
of this  remarkably fast relaxation rate 
of high mass objects could be that the 
``test particle'' approach used 
to calculate two body relaxation time-scales 
fails  for sufficiently  massive particles. 
This is because a heavy mass particle moving in a system
may induce collective effects, altering the density 
around it and changing 
the interaction
strength with neighbouring particles (see Sections 13-14 in part II 
of ref~\protect\cite{Saslaw}).
This effect appears to be amplified in systems evolving from from dynamical 
equilibrium. 
It may be worth noting here that this type of anomalously rapid increase of
high mass particles in violently relaxing objects has  previously been 
reported~\protect\cite{Aarseth and Saslaw,Farouki and Salpeter1}.

\section{The phase transition of enclosed equal mass $N$-body systems}
\label{phatran}

We would now like to know whether the value of the parameter $\mu$ 
at which the phase 
transition between systems that end up as isothermal spheres and 
those that suffer a gravothermal catastrophe takes place is close to 
that given 
by~(\ref{sphex:gravo}), and if this will depend on the initial conditions. 
In the units we employ, $G=M=1$, and the boundary  radius is
also approximately equal to unity so that varying 
$\mu$ basically amounts to changing the total energy.  
If we stick to homogeneous initial spatial configurations, then this will 
amount to
varying the  kinetic energy. We choose this as to obtain 
a virial ratio of
 of $0.69$, $0.6$
$0.5$, $0.4$ and $0.3$. In addition, there was one run started  
in virial equilibrium 
($T/V_{i}$=0.5) but with initial density varying as $1/r^{2}$ instead of 
being homogeneous
(this accordingly decreases the energy). 

We compute the average of $\sigma_{d}$ over
the last hundred outputs (these are produced during the last ten 
crossing times out of a total of usually about a hundred).
Since here we are interested about the final state of the system and not the 
time-scale it 
takes to get there,  the number of particles used in the simulations is not 
very crucial.
In fact, using fewer particles ensures that this state is 
arrived at in a fewer
number of crossing times. Thus one can be more confident that a final state 
has been reached
in a hundred crossing times (say) and use the CPU time saved to be able 
to examine a larger number of initial states.
 The velocities were started with a temperature inversion, 
so that rapid evolution
towards a thermal equilibrium 
 state is expected when such a state exists 
(Section~\ref{slowap}).

The results, which are shown in Table~\ref{sphex:tabT}, 
suggest that the transition between
systems that tend towards thermal equilibria and ones that do not,
  does indeed
take place at a value of $\mu \sim - 0.33$ in accordance with the results of
 Antonov ~\protect\cite{Antonov}
and Lynden Bell and Wood ~\protect\cite{Lynden Bell and Wood}.
 Thus these studies, 
which employed
simple  thermodynamical considerations  applied to ideal gas spheres (and 
were confirmed by local stability analysis in a statistical mechanical 
context~\cite{Horowitz and Katz,Cally and Monaghan}), 
 provide  accurate results concerning the stability of $N$-body
gravitational systems starting from the initial conditions described above. 

However, homogeneous 
spatial initial conditions are obviously, in a sense, closer to the density 
distributions
of isothermal spheres than to collapsed objects (in the sense that these will
have to pass through the less concentrated former state before reaching the
latter).
It was argued~\cite{Kiessling}  on the basis of global minimisation of the 
free energy,
that, for real gravitational systems,
 relation~(\ref{sphex:gravo}) only guarantees 
 {\em local} entropy maxima. Global maxima  being 
collapsed  states with large temperature gradients.
We have therefore examined the approach to equilibrium from spatial 
initial conditions 
where the density decreased according to a power law that is steeper 
than the inverse square
relation associated with an isothermal sphere. 
It was found that for densities decreasing as steeply as $\sim R^{-2.7}$,
the thermal equilibrium final state was recovered. For power laws steeper than that
however, central concentration increased
and the integration stopped (due to numerical difficulties associated
with the fact that the NBODY2 code is not designed to deal with such 
situations~\cite{Aarseth}). 
The details 
(e.g., whether the integration stopped after a few crossing 
times or went on for a while)
depended on the softening and whether the initial 
velocities increased or decreased with radius. However, 
even with very large softening
$\sim 0.1 r_0$ and an initial temperature inversion, a system starting 
with density decreasing
as $R^{-3}$ could not be integrated beyond $84 \tau_c$. 
It may be interesting 
to note here that
the initial velocity distribution 
(with its strong temperature inversion) was
essentially  conserved for
all these tens of crossing times.

It may perhaps be necessary in the future to repeat these simulation with a 
$N$-body
routine that is better suited for integration of the more centrally
 concentrated structures of collapsed gravitational objects 
(e.g., routines involving two body
regularisation) to be completely certain of the following conclusion,
the current evidence however suggests that condition~(\ref{sphex:gravo}) 
indeed only
guarantees that thermal equilibrium states are local entropy maxima.
These correspond to Lane-Embden isothermal spheres which are not true 
isothermal equilibria~(cf., Section~\ref{sphex:genev}. 
Systems starting from certain
initial conditions  may 
still end up
in collapsed states. The condition given in~\cite{Kiessling} for 
this to never
happen  (i.e., for thermal equilibrium states to be global entropy maxima)
is much stricter (with the temperature in units of $k_{B}=1$,
$\mu$ must be  of the order of the inverse of the softening parameter, 
i.e. $\sim + 500$, to prevent collapse). The truly isothermal solutions 
overlap with the Lane-Embden isothermal solutions in this range of $\mu$.

\section{Conclusion}
\label{concs}

In this paper we have studied some aspects of the dynamics of closed 
gravitational systems by means of direct $N$-body simulations. 
Stable $N$-body isothermal spheres are convenient laboratories where 
long term (e.g., ergodic) properties of $N$-body trajectories can be studied 
without 
the distraction caused by the eventual evolution of global properties --- 
which is
inevitable in open systems. 
Up to now this type of study has been conducted
only in the case of one dimensional systems.
The existence of a  final
equilibrium state also means that the relaxation time is well defined. 
Such systems are therefore 
ideal for testing gravitational relaxation theory.
This paper presented an exploratory study where some questions concerning 
the circumstances under which stable 
   $N$-body gravitating isothermal spheres exist
and the relaxation times for evolution towards or away 
from such states were discussed. 
The main findings are as follows.

\begin{enumerate}

\item
It is confirmed that
closed equal mass $N$-body systems that satisfied relation~(\ref{sphex:gravo})
could indeed evolve towards well defined thermal    equilibrium
states, as expected from simple dynamical and thermodynamic considerations 
applied to  systems with a perfect gas equation of state and from local 
stability theory~\cite{BT,Saslaw,Padmanabhan,Horowitz and Katz}.
However, it was found that systems starting from centrally concentrated 
initial states
(e.g., with density decreasing as the inverse cubed of the radius) 
did not evolve
towards thermal equilibrium even when the aforementioned relation 
was satisfied.
This effect was predicted  by Kiessling~\protect\cite{Kiessling} who, 
based on a statistical
mechanical analysis which made used of the minimisation of the free energy,
 suggested that a much more stringent relation than~(\ref{sphex:gravo})
would have  to be satisfied in practice for
all states to tend towards isothermal sphere configurations. 
Unless $\mu$ is very large (of the order of the inverse of the softening
parameter) isothermal spheres are only 
  {\em local} entropy maxima and do not truly represent isothermal 
solutions~(Section~\ref{sphex:genev}).
Whether a particular system will evolve as to end up as an isothermal sphere
will depend on its initial  conditions. 
In particular, local stability of isothermal sphere  configurations 
 will guaranty that some initial states will tend towards such 
equilibria when they exists --- for example states that are initially less
centrally concentrated than the thermal equilibrium configuration. 

\item
The rate at which evolution towards thermal equilibria proceeds depends
a lot on the initial conditions. 
If the system starts sufficiently near a {\em dynamical}
equilibrium state which happens to be different from the true thermal 
equilibrium,
 it will quickly (within a crossing time or so) tend toward the 
dynamical equilibrium. 
The subsequent thermal evolution 
appears to proceed on a time-scale  compatible with slow
 two-body relaxation. 
If, on the other hand,  a system starts from a state  
from where the nearest dynamical 
equilibrium and the true 
thermal equilibrium
 are close (this was found to be the case, for example, for systems which
started from spatially homogeneous states and with initial temperature
inversion), 
it will tend towards this common equilibrium in 
a crossing time 
or so. This can happen even if the conditions for 
violent relaxation are not likely to be  satisfied
--- for example when the system under consideration stays very close to 
virial equilibrium throughout the
evolution and is not clumpy~\protect\cite{Saslaw}. 

\item

The relaxation of anisotropic velocity dispersions was found 
to be three to four times
faster than  energy relaxation (which presumably determines the evolution
towards thermal equilibrium in the case when this evolution is slow). 
One therefore concludes 
that in gravitational systems, different quantities may have different
 relaxation time-scales.
In particular, quantities that depend on the details of a given particle's 
trajectory 
are likely to evolve at a
different rate than its energy, which is a scalar integral of the
 motion~\protect\cite{El-Zant1}.

\item
We have also conducted simulations of 
multimass systems  --- other 
parameters being the same as in the   corresponding single mass runs.
These systems  did not tend to stable thermal equilibria but
 instead underwent gravothermal catastrophe (collapse). 
This appears to be triggered  by the 
statistically irreversible accumulation of heavy mass particles in the 
central areas,  
which changes the ratio of central to boundary densities. 
Since, according to simple dynamical and thermodynamic arguments,
given the  values of the total mass, energy and radius, there exists a 
unique thermodynamically
stable isothermal sphere for a given value of the aforementioned density 
ratio, 
no such equilibria will be possible when central mass is irreversibly 
added while keeping 
other parameters fixed. 

\item
The rate at which the average mass per particle increased in the 
inner $10 \%$ to 
$20 \%$ of the boundary radius  was remarkably fast.
 In particular, it was found that in a situation when there 
was significant departure from virial equilibrium 
(i.e., when violent relaxation may be at work),
mass segregation is observed on surprisingly short time-scales of a few 
crossing 
 times (for systems of 2500 particles). Such effects have 
also been previously 
observed~\protect\cite{Farouki and Salpeter1,Aarseth and Saslaw}.

\end{enumerate} 

It is hoped that this paper has demonstrated some uses of numerical 
experimentation 
with closed $N$-body 
systems and has illustrated that the study of relaxation 
in gravitational systems is not itself a closed subject, 
but is instead a rich and
 largely unexplored area with many potentially interesting phenomena 
yet to be revealed. 

On the theoretical side, an interesting finding is that stable
long lived thermal equilibrium
states can exist which are {\em not} entropy maxima. This 
means that the true entropy maxima (centrally concentrated
 objects with large velocity gradients), 
although compatible with macroscopic constraints 
such as total energy and momentum conservation,
 are not reached --- at least
not over time-scales comparable to the thermal relaxation times.
This of course brings into question the ergodicity of $N$-body 
gravitational systems (over the total energy-momentum subspaces)
and the applicability of the standard postulates of statistical
mechanics to these systems, along with their standard dynamical 
interpretation. 

Two  applications to come out of this effort
that are particularly relevant to the study of stellar dynamics are: 
the evaluation of
the energy relaxation time in cases where a definite thermal 
 equilibrium existed, and
the calculation of the relaxation time of initially 
anisotropic velocity distributions.
In the first case, the relaxation time was found to be compatible with 
that of classical two body
theory. However, it was also compatible  with a recent 
estimate~\protect\cite{El-Zant2}, derived from
the exponential divergence in $N$-body systems in dynamical equilibrium,
 in which the relaxation time scales as 
$\sim \sqrt{N}$.  It is then important to check how the relaxation time 
towards thermal equilibrium
states (which provides the only situation whereas the relaxation time
in a three dimensional gravitational $N$-body system is well defined) 
scales with $N$.
Results from a study of this type are currently 
being analised~\protect\cite{El-Zant and Goodwin}. 

In the second case, we found that the relaxation towards 
isotropic velocity distribution is 
considerably faster than that of energy relaxation. 
If confirmed, this would be
an important result, with astrophysical applications to the study of such 
phenomena as the ellipticity distributions of globular clusters. 
Prigogine and Severne~\protect\cite{Prigogine and Severne}, 
who predicted this effect on basis of a kinetic formulation
of the problem which avoided the introduction of a long range cutoff 
used to eliminate
divergence in the Coulomb logarithm,
estimate that the variation with $N$  of the rate of 
randomisation
``of the kinetic energy 
to that of the
transformation [of the] potential energy into kinetic energy''
 (thermal time-scale of energy relaxation)
goes as $\sim \ln N$. However Prigogine and Severne still retained 
some of simplifying features  
reminiscent of
the original Chandrasekhar formulation of the 
relaxation time: infinite medium 
(i.e., no mean field), weak coupling 
approximation, two body encounters etc.
Obviously then, a direct estimate 
of the variation with $N$ for the aforementioned ratio is also useful.

\begin{table}
\caption{\label{sphex:tabT} 
Values of the relative dispersion in the trace of the velocity dispersion 
tensor
averaged over the last ten crossing times for 500 particle systems. 
Runs normally lasted for a total of a
hundred crossing times except for the run with
 with $E=-0.491$ which was terminated at 33.9 crossing times due to 
numerical difficulties}
\begin{tabular}{ccc}
\multicolumn{3}{c}{ }\\

$T/V_{i}$  &  $E$      &   $\bar{\sigma_{d}}$\\
\hline
0.69       &  - 0.185  &   0.11\\
0.60       &  - 0.238  &   0.11\\
0.50       &  - 0.298  &   0.12\\
0.40       &  - 0.357  &   0.16\\
0.30       &  - 0.417  &   0.22\\
0.50       &  - 0.491  &   0.50\\ 
\end{tabular}
\end{table}


\begin{thebibliography}{}
\bibitem{comamr}
These issues will be discussed in details in~\cite{El-Zant and Goodwin}
\end{thebibliography}
\end{document}